\documentclass[twocolumn]{aastex62}
\pdfoutput=1
\usepackage{graphicx}
\usepackage{bm}
\usepackage{amssymb}
\usepackage{amsmath}
\usepackage{units}
\usepackage{array}
\usepackage{hyperref}
\usepackage{soul}
\usepackage{placeins}
\usepackage{float}

\def\fref{Fig.~\ref}
\def\sref{\S \ref}

\begin{document}

\title{Multi-band light curves from eccentric accreting supermassive black hole binaries}

\correspondingauthor{John Ryan Westernacher-Schneider}
\email{wester5@clemson.edu}

\author[0000-0002-3047-7200]{John Ryan Westernacher-Schneider}
\affiliation{Department of Physics and Astronomy, Clemson University, Clemson, SC 29634, USA}

\author[0000-0002-1895-6516]{Jonathan Zrake}
\affiliation{Department of Physics and Astronomy, Clemson University, Clemson, SC 29634, USA}

\author[0000-0002-0106-9013]{Andrew MacFadyen}
\affiliation{Center for Cosmology and Particle Physics, Physics Department, New York University, New York, NY 10003, USA}

\author[0000-0003-3633-5403]{Zolt\'an Haiman}
\affiliation{Department of Astronomy, Columbia University, New York, NY 10027, USA}

\begin{abstract}
We use long-run, high-resolution hydrodynamics simulations to compute the multi-wavelength light curves (LCs) from thermal disk emission around accreting equal-mass supermassive black hole (BH) binaries, with a focus on revealing binary eccentricity. LCs are obtained by modeling the disk thermodynamics with an adiabatic equation of state, a local blackbody cooling prescription, and corrections to approximate the effects of radiation pressure. We find that modulation of multi-band LCs on the orbital time scale are generally in-phase (to within $\sim\,$2\% of a binary orbital period), but they contain pulse substructure in the time domain that is not necessarily reflected in BH accretion rates $\dot M$. We thus predict that binary-hosting AGN will exhibit highly correlated, in-phase, periodic brightness modulations in their low-energy disk emission. However, detectability of these modulations in multi-wavelength observing campaigns could be seriously compromised because observed stochastic variability in AGNs typically has a higher amplitude than our proposed signal. It is possible that observations over temporal baselines of many binary periods may make the signal more prominent, but this would need to be analyzed carefully. If jet emission is predicted by $\dot{M}$, then we predict a weaker correlation with low-energy disk emission due to the differing sub-peak structure. For the binary parameters we explore, we show that LC variability due to hydrodynamics likely dominates Doppler brightening for all equal-mass binaries with disk Mach numbers $\lesssim 20$. A promising signature of eccentricity is weak or absent ``lump'' periodicity.
We find hints that a significant lag exists between $\dot{M}$ and low-energy disk emission for circular binaries, but they are in-phase for eccentric binaries, which might explain some ``orphan'' blazar flares with no $\gamma$-ray counterpart.
\end{abstract}


\section{Introduction}

Cosmic structure forms hierarchically \citep[][]{white+1978}, thus galaxies merge frequently \citep[e.g.][]{lotz+2011}. Since most galaxies host a supermassive black hole (SMBH) \citep[e.g.][]{kormendy+1995, ferrarese+2005, kormendy+2013}, post-merger galaxies will likely host a supermassive black hole binary (SMBHB) at some stage of their evolution \citep{begelman+1980}. Gravitational waves from orbiting SMBHB systems will likely be observed by the upcoming Laser Interferometer Space Antenna (LISA) mission \citep{LISA2017}, and the NanoGrav collaboration has recently reported a Pulsar Timing Array (PTA) stochastic common-process signal \citep{PTA2020}, which might be the cumulative gravitational wave background sourced by a population of SMBHBs.

For now, the identification of compact SMBHBs (sub-parsec separation, year-like orbital period) relies on electromagnetic (EM) surveys \citep[e.g.][]{graham+2015, charisi+2016, liu+2019, liu+2020, chen+2020}. Those studies, and many others, report evidence for periodically modulated light curves from active galaxies (AGN), ranging from infrared to $\gamma$-ray energies \citep[for a recent review, see][]{derosa+2019}. However, it is not unlikely to observe several cycles of apparent brightness modulation in stochastically variable sources \citep{Vaughan+2016}. Furthermore, binarity is not the only possible cause of genuine periodicity; there are single-SMBH disk processes that might produce quasi-periodic emission as well. These include limit-cycles triggered by Lightman-Eardley instabilities \citep{lightman+1974, FKR}, or iron opacity-driven modulations of the disk thickness \citep{jiang+2020}. Detailed predictions of the multi-band EM signatures of binary accretion are thus motivated to aid in the interpretation of electromagnetic SMBHB candidates.

There are at least two independent causes of periodic variability from accreting binaries: Doppler brightening, and hydrodynamic variability. Doppler modulation is caused by the line-of-sight orbital velocity $v_\parallel$, and induces $\mathcal{O}(v_\parallel/c)$ corrections to the emission from gas around each BH component. This effect can be reliably computed for a range of orbital parameters and orientations \citep[e.g.][]{dorazio1+2015, hu+2020, Charisi2021}. Hydrodynamic variability refers to any other changes in the system's luminous output that are connected to the dynamics of binary accretion. This includes fluctuations in the thermal emission from the disk surfaces arising from adiabatic, viscous, or shock heating. It can also include modulations of a jet luminosity, as induced for example by the time-varying mass accretion rate onto one or both BH components. Our focus in this work is on modeling the hydrodynamic variability of SMBHBs.

Simulated light curves modulated by hydrodynamic processes in accreting equal-mass, circular binaries have been reported previously \citep[e.g.][]{bode+2012, giacomazzo+2012, noble+2012, gold2014, farris+2015b, farris+2015, Tang2018, dAscoli+2018, paschalidis+2021, Gutierrez+2021}. However, eccentricity can evolve due to interaction with the surrounding gas \citep[see e.g.][]{Roedig+2011, Roedig+2014}. \cite{Zrake:2021:eccentric} found that binaries in the gas-driven regime are likely eccentric with $e \simeq 0.4 - 0.5$ \citep[see also][]{DOrazio:2021:eccentric}. Presentations of simulated light curves from eccentric binaries have been limited \citep[see e.g.][]{bogdanovic+2008}. Since gravitational radiation damps eccentricity \citep{peters1964}, only very compact binaries, in the gravitational wave-driven (GW-driven) regime, are expected to be on nearly circular orbits \citep[$e \lesssim 0.01$;][]{Armitage2005, Zrake:2021:eccentric}. It is thus important to predict light curves of both circular and eccentric systems. In particular, robust EM signatures of orbital eccentricity could indicate whether an electromagnetic SMBHB is in the GW- or gas-driven regime, independently of the BH mass and separation estimates.

In this work, we calculate the light curves of accreting SMBHBs with year-like orbital periods, and eccentricity values of $e=0, 0.45$, and $0.7$. Light curves of the thermal disk emission are computed at infrared and optical wavelengths. We also report nominal light curves of non-thermal $\gamma$-ray emission, based on the assumption that the jet luminosity is controlled by the accretion power, which we can accurately measure from our simulations. These predictions can serve as a guide to interpreting AGNs (especially blazars) that exhibit periodicity at different wavelengths.
We pay particular attention to the differentiating characteristics of eccentric versus circular binaries

The paper is organized as follows. In \sref{sec:models} we describe our models for the binary, disk, gas, and cooling prescription, as well as other technical details. In \sref{sec:numerics} we describe pertinent numerical details of our simulations, including the disk initial conditions. Results are presented in \sref{sec:results}. We focus primarily on the following LC observables: modulation periods, amplitudes of hydrodynamic variability, relative power in different electromagnetic bands, and temporal lags and correlations between bands.  We discuss our results in greater detail in \sref{sec:discuss}, including: evidence of binarity (\sref{sec:binevi}), evidence of eccentricity (\sref{sec:eccbinevi}), Doppler brightening (\sref{sec:dop}), considerations which must be made when applying our results to observations, with particular blazars used as a basis for discussion (\sref{sec:sources}), and caveats of our approach (\sref{sec:caveats}). We conclude in \sref{sec:conclude}. The appendix describes numerical prescriptions that were used to attain stable, long-term numerical evolution (\sref{app:fixes}), sensitivity tests and their tabulated results (\sref{app:sens}), details of a Doppler brightening calculation (\sref{app:dop}), and numerical convergence properties of the solution scheme (\sref{app:res}). Throughout this work, ``orbits'' refer to binary orbits unless specified otherwise.

%
%
\section{Models} \label{sec:models}
\subsection{Binaries} \label{sec:binaries}
Our aim is to model the light curves of accreting SMBHB systems in the gas-driven evolutionary phase, with realistic orbital parameters and hydrodynamic conditions. We motivate our fiducial model selections from the following considerations. First, gas accretion tends to equalize the binary component masses \citep[see e.g.][]{Farris+2014,duffel:2020:massratio}, so we have chosen to simulate equal-mass systems. Equilibrium eccentricities for equal-mass binaries have now been measured in simulations, so we choose models with those eccentricity values. Observationally relevant orbital periods for electromagnetic surveys and PTAs are typically year-like, so we choose the component mass and separation accordingly.

The disk hydrodynamic conditions are selected in part to satisfy the requirement that the vast majority of the system's infrared and optical emission is produced on length scales that are resolved in the simulation. In other words, the simulation domain must enclose the part of the circumbinary disk which emits in the infrared, and the thermal emission at unresolved length scales very near the BH components should be mostly at UV and higher energies. This requirement implies the disk surface temperatures must lie in a particular range. The temperature is controlled by an appropriate choice of the disk surface density and effective viscosity. See \sref{sec:sub} \& \sref{sec:sup} for more discussion about how we meet these requirements.

These considerations motivate a fiducial model with mass ratio $q \equiv M_2 / M_1 = 1$, total mass $M\equiv M_1 + M_2 = 8 \times 10^6 M_\odot$, and orbital period $T_{\rm bin} = \unit[1]{yr}$ (semi-major axis $a \simeq \unit[9.7\times10^{-4}]{pc} \simeq 2530 R_g$, where $R_g = GM/c^2$). We consider three different eccentricity values $e \in \lbrace 0, 0.45, 0.7\rbrace$. The circular $e=0$ and eccentric $e=0.45$ cases were found to be equilibrium values for binaries in the gas-driven regime in \cite{Zrake:2021:eccentric}, and a similar equilibrium eccentricity of $e=0.4$ was later reported in \cite{DOrazio:2021:eccentric}. Both studies used a locally isothermal equation of state for the gas, with orbital Mach number $v_{\rm kep}/c_s = 10$, where $v_{\rm kep} = \sqrt{GM/r}$ is the Keplerian orbital velocity and $c_s$ is the isothermal sound speed. 

Calculating thermal emission from the disk surface requires a self-consistent treatment of the gas thermodynamics, so in this study we drop the locally isothermal simplification and solve the hydrodynamics equations with an adiabatic equation of state (see \sref{sec:gas}) and radiative cooling prescription (see \sref{sec:cooling}). As demonstrated in \cite{Tiede2020}, the disk thermodynamics can have a significant effect on the binary orbital evolution, so the equilibrium value $e \simeq 0.45$ is expected to be approximate in our case, and we leave to future work a determination of the equilibrium value of eccentricity with more realistic thermodynamics. A more extreme eccentricity of $e=0.7$ is included in our study, in order to check how generic our results are.

Embedded in a thin accretion disk, these binaries are likely in the gas-driven regime of orbital evolution, where significant eccentricity is expected. We estimate the semi-major axis at which there is a transition between gas-driven and GW-driven regimes, $a_{\rm GW}$,
by equating the rate of gas-driven inspiral to the rate of GW-driven inspiral. This is done by plugging Post-Newtonian evolution \citep[][]{peters1964} of the semi-major axis $a$ into the following relation and then solving for $a$:
\begin{eqnarray}
    \frac{da}{dM} = -\ell \frac{a}{M},
\end{eqnarray}
where $\ell=\mathcal{O}(1)$ is an ``eigenvalue'' \citep{paczynski1991, popham+1991} determined by gas accretion physics. For example, an effective value of $\ell\simeq 0.43$ was reported for binaries with non-zero, near-equilibrium eccentricity \citep{DOrazio:2021:eccentric}. The result of this substitution is
\begin{eqnarray}
    a_{\rm GW}^4 = \frac{M}{\dot{M}} \frac{1}{\ell} \frac{64}{5} \frac{G^3 M_1 M_2 M}{c^5 (1-e^2)^{7/2}} \left( 1 + \frac{73}{24}e^2 + \frac{37}{96}e^4 \right). \label{eq:agw}
\end{eqnarray}
Specializing to equal-mass binaries with eccentricities of either $e=0$ or $e=0.45$, and scaling Eq.~\eqref{eq:agw} according to our target system parameters (described fully in subsequent sections), we obtain the expression for arbitrary total binary mass $M$, accretion rate $\dot{M}$, and accretion eigenvalue $\ell$, relative to our fiducial binary:
\begin{eqnarray}
    a_{\rm GW} &\simeq& \unit[10^{-3}]{pc} \times B(e) \\ &\times& \ell^{-1/4} \left(\frac{\dot{M}}{10\dot{M}_{\rm Edd}} \right)^{-1/4} \left( \frac{M}{8 \times 10^6 M_\odot} \right)^{3/4} \nonumber
\end{eqnarray}
where $B(e=0) \simeq 0.73$ and $B(e=0.45)\simeq 1$. In other words, our fiducial binary is very close to having gravitational waves start to dominate over gas torques.

\subsection{Disk} \label{sec:disk}
When discussing the general characteristics of our target system, we speak of a single black hole of mass $M$, surrounded by a geometrically thin and optically thick Shakura-Sunyaev accretion disk model with constant-$\alpha$ viscosity~\citep{SS1973} ($\alpha=0.1$) undergoing near-Keplerian rotation. However, we will be placing a binary in the system instead of a single black hole. We largely follow the purely Newtonian treatment given in~\cite{Goodman2003}, except we relate the disk effective temperature $T_{\rm eff}$ to the mid-plane temperature $T$ via \begin{eqnarray}
T_{\rm eff}^4 = \frac{4}{3} \frac{T^4}{\kappa \Sigma},
\end{eqnarray}
and we use the sound speed appropriate for a fluid composed of a nontrivial mixture of gas and radiation.\footnote{Rather than the relation $T_{\rm eff}^4 = 2T^4/(\kappa\Sigma)$
and isothermal sound speed $c_s=\sqrt{P/\rho}$ used in~\cite{Goodman2003}. The sound speed for a mixture of gas and radiation pressure is given by $c_s^2 = \gamma_\beta \mathcal{P}/\Sigma$, where $\gamma_\beta \equiv \beta + (4-3\beta)^2(\Gamma -1)/(\beta + 12 (\Gamma-1)(1-\beta))$, $\Sigma$ and $\mathcal{P}$ are the vertically integrated mass density and total pressure, $\beta$ is the gas pressure fraction $\beta\equiv \mathcal{P}_{\rm gas}/\mathcal{P}$, and $\Gamma$ is the adiabatic index of the gas component of the fluid. $\gamma_\beta$ interpolates between $\gamma_0=4/3$ and $\gamma_1=\Gamma$.}

We assume the black hole accretes at $10\times$ the Eddington rate, i.e.~$\dot{M} = 10\, \dot{M}_{\rm Edd}$, where $\dot{M}_{\rm Edd} = L_{\rm Edd}/(\eta c^2)$ and the radiative efficiency is assumed to be $\eta=0.1$. This choice of accretion rate is primarily motivated by obtaining a numerically tractable Mach number $\mathcal{M} \sim \mathcal{O}(10)$, which also allows a comparison with past work.

The disk is Toomre-stable out to a radius such that $Q\geq c_s \Omega/(\pi G \Sigma) = 1$, where $\Omega$ is the Keplerian angular frequency of the gas and $\Sigma = 2 h \rho$ is the surface density. The disk semi-thickness $h$ around a single black hole is given by an approximate solution to the equation of vertical hydrostatic balance, $h\simeq \sqrt{P/\rho}\, \Omega^{-1}$, and $P$ is the total (i.e.~gas and radiation) pressure. In terms of the semi-major axis of our chosen binary, $a\simeq \unit[10^{-3}]{pc}$, the disk is Toomre-stable out to $r\simeq 11\, a$. We neglect the self-gravity of the disk, which is justified to the extent that the disk is Toomre-stable out to a radius significantly larger than $a$.

The orbital Mach number profile $\mathcal{M}(r)=v_{\rm K}(r)/c_s(r)$ increases rapidly with radius. For example, $\mathcal{M}(a) \simeq 7$, $\mathcal{M}(1.5a) \simeq 11$, and $\mathcal{M}(3a) \simeq 21$. (Note that in our simulations, the Mach number profile develops self-consistently from a balance of heating and cooling.) The effective optical depth $\tau_{\rm eff} = \sqrt{\tau_{\rm absorp} (\tau_{\rm absorp} + \tau_{\rm scattering})}$ at some radii of interest from the single black hole is $\tau_{\rm eff}(r\!=\!a) \simeq 10^5$ and $\tau_{\rm eff}(r\!=\!0.02 a) \simeq 2$, where we estimate $\tau_{\rm absorp}$ from the Planck mean opacities tabulated online\footnote{\url{https://aphysics2.lanl.gov/apps/}} for Milky Way elemental abundance, and we take the scattering opacity to be due to electron scattering. Our assumption of optical thickness and blackbody cooling spectra are justified to the extent that $\tau_{\rm eff} > 1$. The viscous time scale $t_\nu \equiv (2/3) r^2/\nu$ becomes equal to the cooling time scale $t_{\rm cool} \equiv U/\dot{Q}$ at $r\simeq 0.18a \simeq \, 460 R_g$, where $U$ is the surface density of internal energy and $\dot{Q}$ is the cooling rate per area. Our assumption of radiative efficiency is justified to the extent that $t_{\rm cool} < t_\nu$, which may be violated in the innermost regions of the minidisks in our binary simulations.

In our target model above, radiation pressure dominates the disk. For example, the gas pressure fraction $\beta \equiv P_{\rm gas}/P$ at some radii of interest are $\beta(r\!=\!a) \simeq 9.4\times10^{-4}$ and $\beta(r\!=\!3a) \simeq 0.016$. Including radiation pressure in simulations is a nontrivial task algorithmically, and the disk may be subject to limit-cycle instabilities~\citep{lightman+1974,FKR}. Thus, we use only gas pressure in this work, and below we describe our strategy to approximate the effects of radiation pressure.

\subsection{Gas} \label{sec:gas}
We use a $\Gamma$-law equation of state with $\Gamma=5/3$, yielding the equation of state $\mathcal{P} = \Sigma \epsilon (\Gamma -1)$, where $\Sigma$ and $\mathcal{P}$ are respectively the vertically-integrated mass density and pressure, and $\epsilon$ is the specific internal energy density at the mid-plane of the disk. We use constant-$\alpha$ viscosity yielding a kinematic shear viscosity $\nu = \alpha c_s h$, where $c_s^2 = \Gamma \mathcal{P}/\Sigma$. For a binary, the disk semi-thickness is $h = \sqrt{\mathcal{P}/\Sigma}/\tilde{\Omega}$, where $\tilde{\Omega} = \sqrt{GM_1/r_1^3 + GM_2/r_2^3}$ and $r_1$, $r_2$ are the distances from a field point to the respective point masses $M_1$, $M_2$.

The vertically-integrated Newtonian fluid equations keep the lowest nontrivial order in powers of $z/r$ under the conditions of a thin disk ($h/r\ll 1$) and mirror symmetry about $z=0$. These equations read
\begin{eqnarray}
    \partial_t \Sigma + \nabla_j \left( \Sigma v^j \right) &=& S_{\Sigma} \label{eq:mass} \\
    \partial_t \left( \Sigma v_i \right) + \nabla_j \left( \Sigma v^j v_i + \delta^j_i \mathcal{P} \right) &=& g_i + \nabla_j \tau^j_i + S_{p, i} \label{eq:mom} \\
    \partial_t E + \nabla_j \left[ \left( E+\mathcal{P} \right) v^j \right] &=& v^jg_j + \nabla_j \left( v^i \tau^j_i \right) \nonumber\\
    &-& \dot{Q} + S_{E} \label{eq:en},
\end{eqnarray}
where $v^i$ is the mid-plane horizontal fluid velocity; $E=\Sigma \epsilon + (1/2)\Sigma v^2$ is the vertically-integrated energy density; $g_i$ is the vertically-integrated gravitational force density; $\tau^j_i = \Sigma \nu \left( \nabla_i v^j + \nabla^j v_i - (2/3)\delta^j_i \nabla_k v^k\right)$ is the viscous stress tensor (in a form that is trace-free in a 3-dimensional sense)\footnote{Note this viscous stress tensor should be understood as being inserted \emph{after} vertical integration of the perfect fluid equations, as a model of unresolved turbulence and magnetic fields.} with zero bulk viscosity; $S_{\Sigma}$, $S_{p,i}$, and $S_E$ are the mass, momentum, and energy sinks; and $\dot{Q}$ is a radiative cooling term, described in \sref{sec:cooling}. Thermal conductivity is neglected.

\subsection{Gravity}
We model the vertically-integrated gravitational force from a point mass $M_n$ as that arising from a Plummer potential,
\begin{eqnarray}
    \Phi_n = -\frac{G M_n}{\sqrt{r_n^2 + r_s^2}} \label{eq:plummer},
\end{eqnarray}
where $r_n$ is the distance from a field point to the $n$th mass, and $r_s$ is the softening length. In this work, we set the softening length equal to the sink radius, $r_s=r_{\rm sink}$ (defined in \sref{sec:sinks}). Alternative models of the vertically-integrated gravitational force, such as assuming $r_s \propto h$, may yield stronger gravity near point masses, which may alter the gas dynamics appreciably. We leave a careful study of this to future work.

\subsection{Sinks} \label{sec:sinks}

To model accretion onto scales below the grid separation, we use torque-free sink prescriptions \citep{dempsey+2020, Dittmann+2021} for each point mass. The torque-free sink models a steady accretion flow with a torque-free inner boundary, implying that angular momentum is advected inward and viscously transported outward in equal measure. The sink terms are
\begin{eqnarray}
    S_{\Sigma} &=& - s \Omega \Sigma \sum_n w_n \label{eq:msink}\\
    S_{p, i} &=& - s  \Omega \Sigma \sum_n v^*_{i,n} w_n \label{eq:psink}\\
    S_E &=& - s \Omega \sum_n E^*_n w_n, \label{eq:Esink}
\end{eqnarray}
where the star superscript is defined below, $s$ is a dimensionless sink rate, $\Omega=\sqrt{GM/a^3}$ is the Keplerian angular frequency of the binary, and $w_n$ is a dimensionless window function defined in terms of a sink radius $r_{\rm sink}$ and a distance $r_n = \sqrt{(x-x_n)^2 + (y-y_n)^2}$ from a field point $(x,y)$ to the $n$th point mass $(x_n,y_n)$ as
\begin{eqnarray}
    w_n = \exp{\lbrace -(r_n/r_{\rm sink})^4 \rbrace}.
\end{eqnarray}
For $r_n/r_{\rm sink} > 4$, we truncate the window function to $w_n=0$. Torque-free sinks are achieved through an adjustment of the velocities, which appear in Eqs.~\eqref{eq:psink} \&~\eqref{eq:Esink} with a star superscript: 
\begin{eqnarray}
    \vec{v}^*_{n} \equiv \left((\vec{v} - \vec{v}_{M_n})\cdot \hat{r}_{M_n}\right) \hat{r}_{M_n} + \vec{v}_{M_n},
\end{eqnarray}
 where $v_{i,M_n}$ is the velocity of point mass $M_n$ and $\hat{r}_{M_n}$ is the unit radial vector in a coordinate system centered on point mass $M_n$. This adjustment removes the angular component of the velocity in the frame that moves with and is centered on the point mass. When used in Eqs.~\eqref{eq:psink} \&~\eqref{eq:Esink}, torque-free sinks are attained. Note that the kinetic energy in Eq.~\eqref{eq:Esink} has been modified such that $E^*_n \equiv \Sigma \epsilon + (1/2) \Sigma \,(\vec{v}^{\,*}_n)^2$.

\subsection{Accretion rate} \label{sec:accrate}

In our target system, we assume an accretion rate of $10\times$ the Eddington rate in the background disk with radiative efficiency $\eta=0.1$. Recent numerical work has shown that such super-Eddington accretion rates can be physically realized \citep[see e.g.][]{Jiang+2019}, but our motivation for this choice is primarily to obtain numerically tractable Mach numbers. However, note that the rate of mass flux in the disk is not necessarily the rate of mass absorbed by the black hole, since there may be outflows which occur at subgrid scales or are otherwise uncaptured phenomena in our simulations. The surface density and viscosity are related to the accretion rate via $\dot{M} = 3\pi \Sigma \nu$ \citep[see e.g.][]{FKR}.

\begin{figure} 
\centering
\includegraphics[width=0.47\textwidth]{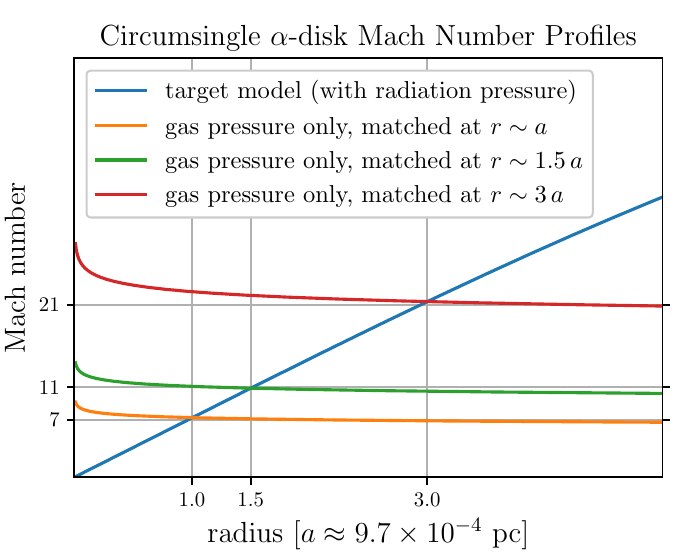}
\caption{Mach number profiles of circumsingle $\alpha$-disks, which are used as initial conditions for binary simulations. The Mach number in models with only gas pressure are matched to the radiation-dominated target model at specific radii.} \label{fig:initialmach}
\end{figure}

As mentioned above, we are modeling our target system using a gas-dominated fluid. Our strategy to do so is to match the Mach number of the initial disk profile of our target system at one particular radius, via large adjustments of the accretion rate into the extremely super-Eddington regime. The extremely super-Eddington accretion rate should be viewed as an artificial aspect of our gas-dominated simulations, whose purpose is only to yield Mach numbers in a similar range as the radiation-dominated target model. As a check of the sensitivity of our results, in our gas-dominated model we study three cases where the initial Mach number is matched to the target system at radii $r=\lbrace a, 1.5a, 3a \rbrace$ (where the initial Mach number is $\mathcal{M} \simeq \lbrace 7, 11, 21 \rbrace$, respectively). The accretion rates which achieve these Mach numbers in the gas-dominated model are respectively $A\equiv \lbrace 6.7\!\times\!10^6,\, 8.15\!\times\!10^5,\, 2.3\!\times\!10^4 \rbrace$ times the Eddington rate. See \fref{fig:initialmach} for a comparison of the initial Mach profiles of the target model and its gas pressure approximations.

Since the effective temperature is related to the accretion rate via
\begin{eqnarray}
    T_{\rm eff}^4 \propto \dot{M},
\end{eqnarray}
these artificially high accretion rates in the gas-dominated models result in artificially high effective temperatures, which affect the light curves in different bands. Thus, when computing light curves in post-processing, we adjust the effective temperature uniformly back down to our target system via the map $T_{\rm eff}^4 \rightarrow T^4_{\rm eff}\times 10/A$, where the factor of $10$ comes from the target system's accretion rate of $10\times \dot{M}_{\rm Edd}$. In a steady disk, this mapping exactly reproduces the effective temperature profile of our target system; in the presence of a binary, there is nontrivial dynamics, so this mapping is approximate. In our binary simulations, the artificially high accretion rates scale up viscous heating, radiative cooling, and shock heating in roughly the same proportions. To see this, note that viscous heating scales linearly with $\dot{M}$ as long as velocity profiles remain approximately invariant,
\begin{eqnarray}
    \nabla_j\left(v^j\tau^j_i\right) \propto \Sigma\nu \propto \dot{M}.
\end{eqnarray}
Radiative cooling scales similarly,
\begin{eqnarray}
    \dot{Q} \propto T_{\rm eff}^4 \propto \dot{M}.
\end{eqnarray}
Lastly, shock heating is described by the Rankine-Hugoniot condition for the energy equation, which scales linearly with $\dot{M}$ as long as the adiabatic index $\Gamma$ and Mach number $\mathcal{M}$ are also held approximately fixed:
\begin{eqnarray}
    E &=& \Sigma \epsilon + (1/2)\Sigma v^2 \nonumber\\
    &=& \mathcal{P} \left( \frac{1}{\Gamma-1} + (1/2)\mathcal{M}^2\Gamma \right) \nonumber\\
    &\propto& \mathcal{P} \propto \Sigma c_s h \propto \Sigma\nu \propto \dot{M}.
\end{eqnarray}
We believe the largest discrepancy will be in shock heating, primarily because the Mach number is not matched well everywhere (the radiation-dominated model has a steep increase of $\mathcal{M}$ with $r$, whereas the gas-dominated model has a much flatter profile -- see \fref{fig:initialmach}).

\subsection{Cooling} \label{sec:cooling}
We model radiative cooling assuming geometrically thin, optically thick gas, using 
\begin{eqnarray}\label{eq:cooling}
    \dot{Q} = \frac{8}{3} \frac{\sigma T^4}{\kappa \Sigma},
\end{eqnarray}
where $\sigma$ is the Stefan-Boltzmann constant, $\kappa=0.4$ cm$^{2}$ 
g$^{-1}$ is the opacity due to electron scattering, and $T=(m_{\rm p}/k_{\rm B}) \mathcal{P}/\Sigma$ is the mid-plane temperature assuming hydrogen dominates the gas density. In post-processing, the effective temperature on one face of the disk is obtained from $\dot{Q}$ via the relation
\begin{eqnarray}
    \dot{Q} = 2\sigma T_{\rm eff}^4,
\end{eqnarray}
where the factor of $2$ comes from the fact that cooling occurs on both faces of the disk.

\subsection{Spectra}
We assume blackbody emission from each cell of our domain, which allows us to compute the cooling luminosity in different bands. We therefore neglect possible dynamical variations in optical thickness. We neglect Doppler effects, so our light curves are valid for observers who are oriented face-on to the disk. We take the infrared range of wavelengths $\lambda$ to be $700\,$nm$-1\,$mm, and the optical range to be $400\,$nm$-700\,$nm.

\subsection{Sub-sink emission} \label{sec:sub}
To estimate the cooling luminosity occurring below the sink scale, we model the sub-sink disks as gas-dominated Newtonian multi-color $\alpha$-disks with $\Gamma=5/3$ that is instantaneously in equilibrium with the respective accretion rates as registered by the sink terms. We integrate the cooling luminosity from the sink radius $r_{\rm sink}$ down to the innermost stable orbit $r=6 GM_n/c^2$ of an assumed Schwarzschild black hole. The purpose of this estimate is to gauge how much of the electromagnetic emission is resolved on the grid. We report the infrared and optical bands since we can resolve them to a high degree, whereas only a small minority of UV and higher-energy bands are resolved. Note that we perform the effective temperature mapping described in \sref{sec:accrate} when estimating the sub-sink emission.

\subsection{Super-domain emission} \label{sec:sup}
Similarly to our estimate of the sub-sink emission, we also estimate the missing emission at radii beyond our computational domain, except the assumed accretion rate is the constant one prescribed in the initial conditions. Super-domain optical emission is negligible ($\mathcal{O}(10^{37})\,$erg/s), whereas infrared emission is not. A caveat is that the super-domain emission may not be well-represented by an axisymmetric disk, since it may be Toomre-unstable or subject to ionization instabilities.

%
%
\section{Numerics} \label{sec:numerics}
We use Cartesian coordinates on a square domain with side length $2 D$, where $D=15 a$ is the fiducial domain radius measured from the center of the grid. Our fiducial resolution is $\Delta x = \Delta y = 0.01 a$, and we use a Courant-Friedrichs-Lewy factor in the range $C\equiv 0.02-0.1$ (depending on how demanding the simulation proves to be), giving a time step $\Delta t = C \Delta x / {\rm max}(|v_x|+c_s,\, |v_y|+c_s)$. We utilize the Harten-Lax-van Leer-Einfeldt (HLLE) flux formula, piecewise-linear extrapolation of primitive variables to the cell interfaces, and 2nd-order Total Variation Diminishing (TVD) Runge-Kutta time stepping. Slope limiting is done using the generalized minmod limiter, with parameter $\theta = 1.5$. This value yields a good balance between robustness and low numerical diffusion.

The cooling term in Eq. \ref{eq:cooling} is included on the right-hand-side of the energy evolution equation, Eq. \ref{eq:en}. In regions where the cooling time scale is shorter than the time step size $\Delta t$, the gas temperature can go negative unless additional care is taken. A robust approach, described in \cite{Ryan+2017} and which we have adopted in our code, is to apply the cooling term in a semi-implicit manner, where the internal energy subtracted in a time step is determined by analytic integration of the cooling curve over the time interval $\Delta t$, such that $\Delta Q = \int_t^{t + \Delta t} \dot Q(t') dt'$. This procedure is effective, and not costly in terms of performance.

``Buffer'' source terms are employed in the vicinity of the grid boundaries which drive the solution to the initial conditions for the disk. This results in a squishy outer boundary which prevents the square grid edges from propagating artifacts into the inner region of the domain. For fluid variables $\vec{U}$ and initial condition $\vec{U}_0$, the buffer source terms have the form
\begin{eqnarray}
    \vec{B} \equiv -f(r) \Omega\vert_{r\!=\!D} \left( \vec{U} - \vec{U}_0\right)\vert_{r\!=\!D-0.1a},
\end{eqnarray}
where $f(r)$ increases linearly from 0 at $r\!=\!D\!-\!0.1a$ to 1000 at $r\!=\!D$ (and is zero otherwise), and $\Omega|_{r\!=\!D}$ is the Keplerian angular frequency at the domain radius.

Additional artificial prescriptions for code stability are described in Appendix~\ref{app:fixes}. Sensitivity tests of our science results to various prescriptions are described in Appendix~\ref{app:sens}. Where applicable, we quote results with uncertainties as indicated by the sensitivity tests.

\subsection{Initial conditions} \label{sec:init}
The disk initial conditions are
\begin{eqnarray}
    \Sigma &=& \Sigma_0 \left(\frac{r_{\rm soft}}{a}\right)^{-3/5} \nonumber\\
    \mathcal{P} &=& \mathcal{P}_0 \left(\frac{r_{\rm soft}}{a}\right)^{-3/2} \nonumber\\
    \vec{v} &=& \sqrt{\frac{GM}{r_{\rm soft}}} \, \hat{\phi},
\end{eqnarray}
where $r_{\rm soft} = \sqrt{r^2 + r_s^2}$. We also initialize a central cavity of radius $2a$ with a sharp edge by multiplying $\Sigma$ and $\mathcal{P}$ by 
\begin{eqnarray}
    10^{-4} + (1-10^{-4})\exp\lbrace -(2a/r_{\rm soft})^{30} \rbrace.
\end{eqnarray}
For the case $\mathcal{M}(1.5 a)\simeq 11$, we set $\Sigma_0 \simeq 0.48 M/a^2$, $\mathcal{P}_0 \simeq 0.002 M\Omega_{\rm bin}^2$. For the case $\mathcal{M}(a)\simeq 7$, we set $\Sigma_0 \simeq 1.7 M/a^2$, $\mathcal{P}_0 \simeq 0.019 M\Omega_{\rm bin}^2$. For the case $\mathcal{M}(3a)\simeq 21$, we set $\Sigma_0 \simeq 0.057 M/a^2$, $\mathcal{P}_0 \simeq 6.7\!\times\!10^{-5} M\Omega_{\rm bin}^2$.

\subsection{Grid refinement}
We evolve all runs initially with a resolution of $\Delta x = 0.02 a$ and sink radius $r_{\rm sink} = 2 \Delta x$. The disk settles into a statistically quasi-steady state on the order of a viscous time, which for our different Mach number runs is $t_\nu \simeq 50 - 500$ orbits. When the runs are near our desired analysis time (typically 600 orbits), we refine the grid uniformly to a resolution of $\Delta x = 0.01 a$, using zeroth-order interpolation and keeping the sink size fixed (so that $r_{\rm sink} = 4 \Delta x$ following refinement), and run for an additional 100 orbits in order to allow the system to settle (typically to 700 orbits). Based on qualitative inspection of the mass accretion rates, the system typically settles within $\lesssim 10$ orbits following refinement (and more quickly for higher resolution and smaller sink radius). The duration for our analysis is the subsequent 100 orbits (typically orbits $700-800$). We present tests of sensitivity to the analysis time and the duration of pre-analysis evolution at $\Delta x = 0.01a$ in Appendix \sref{app:sens} (test labels: AT and ET, respectively).

In our resolution and sink-shrinking tests (test labels: $\Delta x$ and $r_{\rm sink}$, respectively), we refine the grid again at 700 orbits from $\Delta x = 0.01 a$ to $\Delta x = 0.005a$, and evolve at the latter resolution until 800 orbits. In the case of the sink-shrinking test, we keep the ratio $r_{\rm sink}/\Delta x=4$ fixed, since we believe the sink is well-resolved at this value, and we scale up the dimensionless sink rate according to the viscous time at the sink radius, $s \propto r_{\rm sink}^2$. The duration for our analysis at a resolution of $\Delta x = 0.005a$ is then roughly orbits $800-840$.

In our sink-shrinking test, we refine the grid once more at 800 orbits to $\Delta x = 0.0025 a$, once again keeping the ratio $r_{\rm sink}/\Delta x = 4$ fixed and scaling up the sink rate according to $s\propto r_{\rm sink}^2$. Since the computational cost of running at this resolution is so high, we begin analysis immediately, with the analyzed duration being roughly orbits $800-814$.

\subsection{Integrating Planck spectra}
The blackbody luminosity from an area element $dA$ of the computational domain between frequencies $\nu_1$ and $\nu_2$ is
\begin{eqnarray}
    dL = \pi dA \int_{\nu_1}^{\nu_2} \frac{2 h \nu^3 c^{-2} d\nu}{\exp\{\frac{h\nu}{kT_{\rm eff}}\} - 1}. \label{eq:cell_lum}
\end{eqnarray}
For computational expedience, we perform the frequency integral in Eq.~\eqref{eq:cell_lum} approximately using the method of~\cite{integrateplanck}, with the sum in their equation (6) carried out to $n=15$, which resulted in $\mathcal{O}(1)\%$ accuracy in our tests. When integrating Eq.~\eqref{eq:cell_lum} over the area of the disk, we omit the buffer region $r\!>\!D\!-\!0.1a$.

\begin{figure} 
\centering
\includegraphics[width=0.47\textwidth]{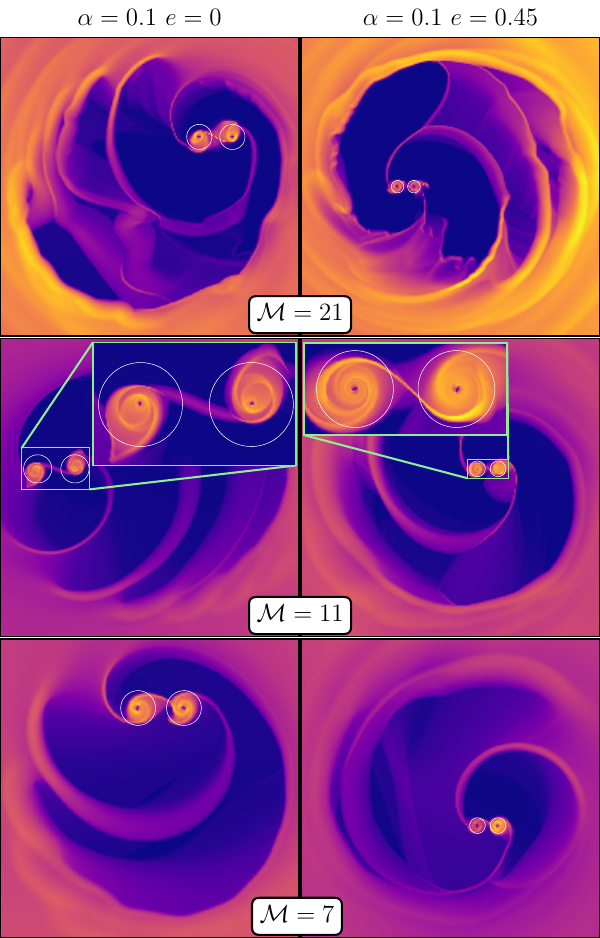}
\caption{Surface density snapshots, raised to the $1/4$th power to improve contrast. Left and right columns are circular and eccentric binaries. All binaries are orbiting counter-clockwise. Eccentric binaries are shown at pericenter. From top to bottom, rows correspond to three Mach number runs which we have labeled $\mathcal{M}\in\lbrace 21,11,7 \rbrace$. The $\mathcal{M}=11$ case displayed is our highest resolution run, with $\Delta x=0.0025 a$ and $r_{\rm sink}=4\Delta x$. The other Mach numbers are at $\Delta x=0.01 a$ and $r_{\rm sink}=4\Delta x$. White circles are centered on each black hole, with radius $r_{\rm Egg}\equiv a\times 0.49/(0.6+\ln(2))$ \citep{eggleton1983} for the circular binary and $(1-e)\times r_{\rm Egg}$ for the eccentric binary. In physical units, these radii are $\simeq 3.7\times10^{-4}\,$pc for the circular binary and $\simeq 2.0\times10^{-4}\,$pc for the eccentric one.} \label{fig:snaps}
\end{figure}
%
%
\section{Results} \label{sec:results}

\begin{figure*}[ht]
\centering
\includegraphics[width=1\textwidth]{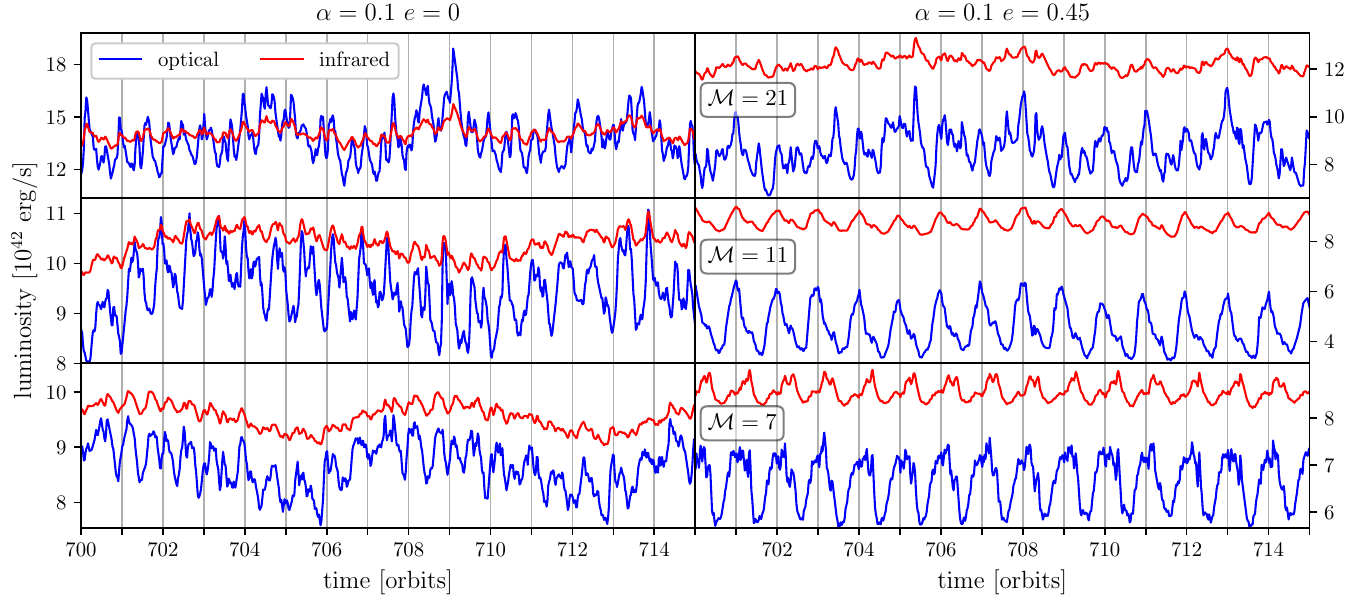}
\caption{Light curves in the optical (blue) and infrared (red) bands for circular ($e=0$, left column) and eccentric ($e=0.45$, right column) binaries. From top to bottom, rows correspond to Mach numbers 21, 11, 7. The lump period features prominently in the circular case. Each light curve includes corrections from our estimates of sub-sink and super-domain emission.} \label{fig:LCMach}
\end{figure*}

\begin{figure*}[ht]
\centering
\includegraphics[width=1\textwidth]{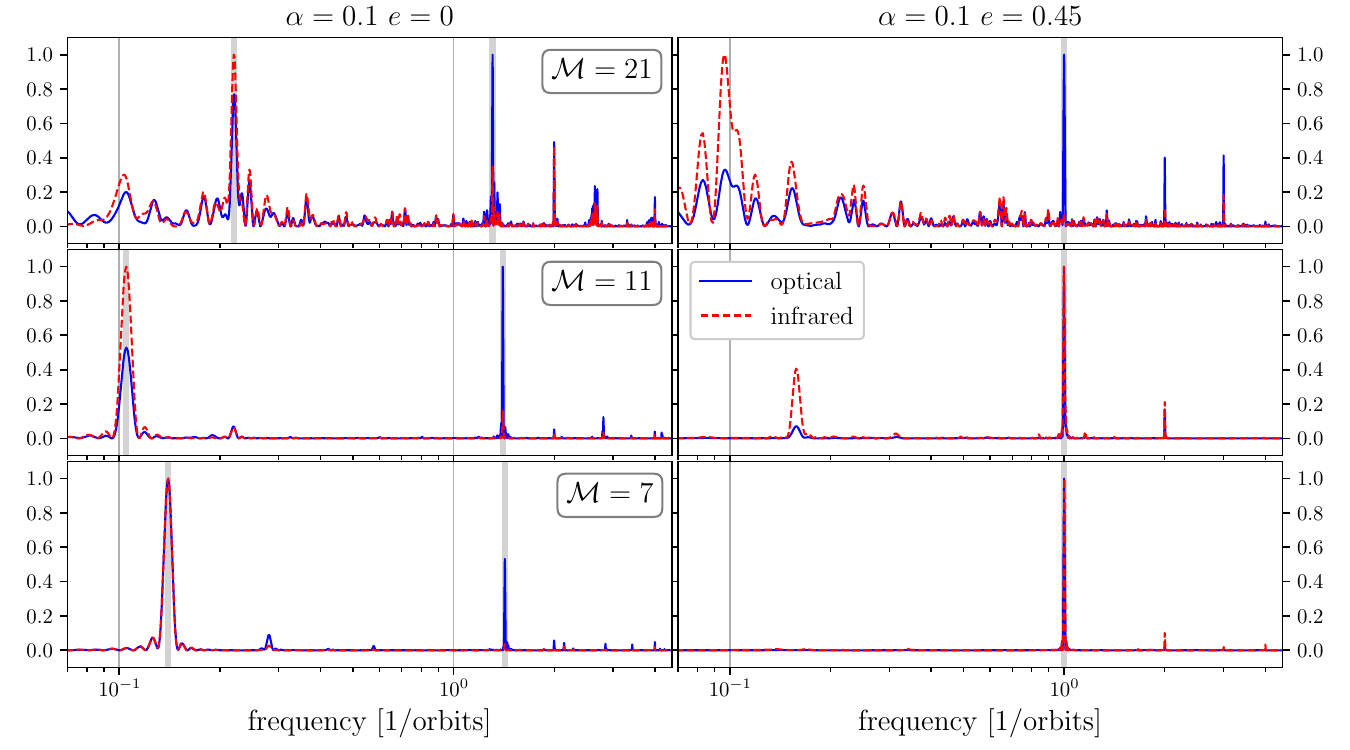}
\caption{Normalized Lomb-Scargle periodograms of the optical (blue) and infrared (red) light curves for circular ($e=0$, left column) and eccentric ($e=0.45$, right column) binaries. From top to bottom, rows correspond to Mach numbers 21, 11, 7. The peak in the circular case at $\simeq 1.4\,$orbits$^{-1}$ decreases towards the orbital frequency in our sink-shrinking tests (see \fref{fig:ftconverge}).} \label{fig:specMach}
\end{figure*}

\begin{figure} 
\centering
\includegraphics[width=0.47\textwidth]{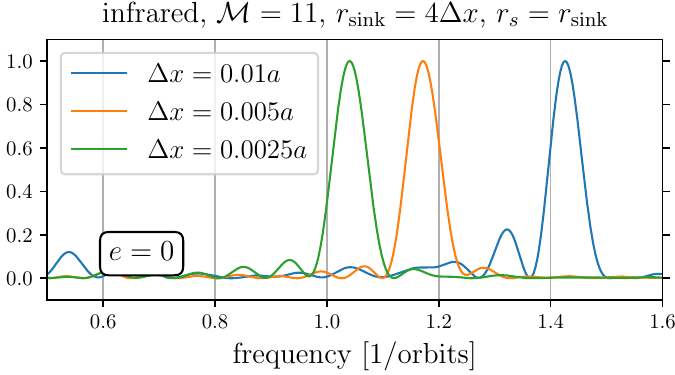}
\caption{Normalized Lomb-Scargle periodograms of the infrared light curves for the $\mathcal{M}\simeq 11$ circular binary at different sink sizes. Resolution is increased as the sink size is reduced, such that the sink radius is always $r_{\rm sink}=4\Delta x$. The 1.4 orbits$^{-1}$ frequency appears to approach the orbital frequency in the limit of small sink. Temporal baselines have been truncated to the length of the highest resolution case, $\sim 14$ orbits, which has widened the peaks to a similar width.} \label{fig:ftconverge}
\end{figure}

\begin{figure}[ht]
\centering
\includegraphics[width=0.47\textwidth]{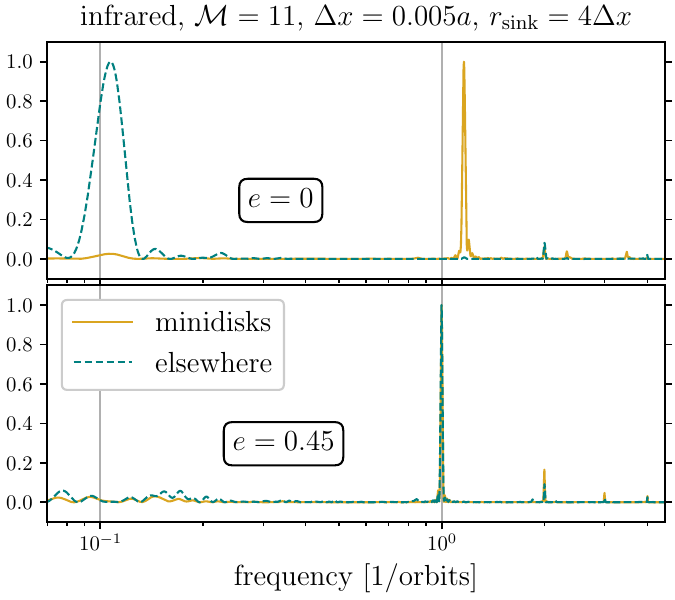}
\caption{Normalized Lomb-Scargle periodograms for the total minidisk emission and emission coming from elsewhere (primarily accretion streams and the cavity wall). As a representative case, we only show the infrared emission from the run with $\mathcal{M}=11$, $\Delta x = 0.005a$, $r_{\rm sink}=4\Delta x$. Optical emission (not shown) is similar, except for a slightly less prominent lump periodicity. For the circular binary, the lump periodicity is almost absent in the minidisk emission. Emission from the accretion streams and the cavity wall has a noticeable peak at $2\,$ orbits$^{-1}$, which we interpret as ejected accretion streams shock-heating the cavity wall twice per orbit.} \label{fig:ftMD}
\end{figure}

\begin{figure} 
\centering
\includegraphics[width=0.47\textwidth]{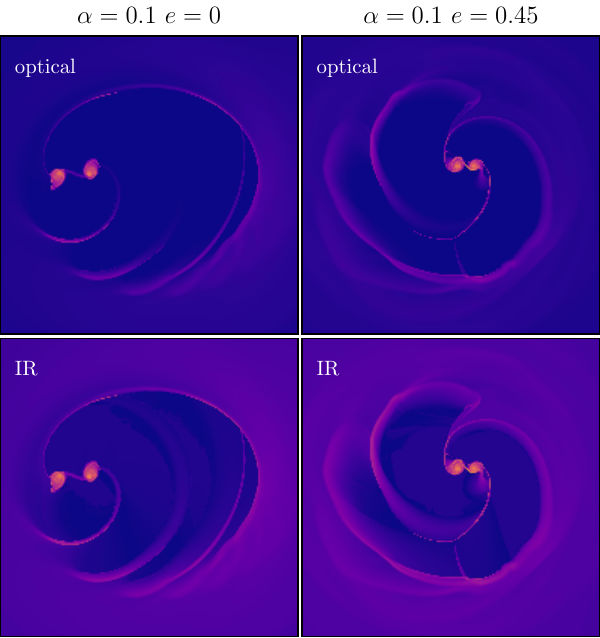}
\caption{Luminosity map snapshots, raised to the $1/4$th power to improve contrast. Only the $\mathcal{M}=11$ case is shown, from our highest resolution runs. Left and right columns are circular and eccentric binaries. All binaries are orbiting counter-clockwise. Eccentric binaries are shown at pericenter. Top and bottom rows optical and infrared maps.} \label{fig:lummaps}
\end{figure}

\begin{figure*}[ht]
\centering
\includegraphics[width=1\textwidth]{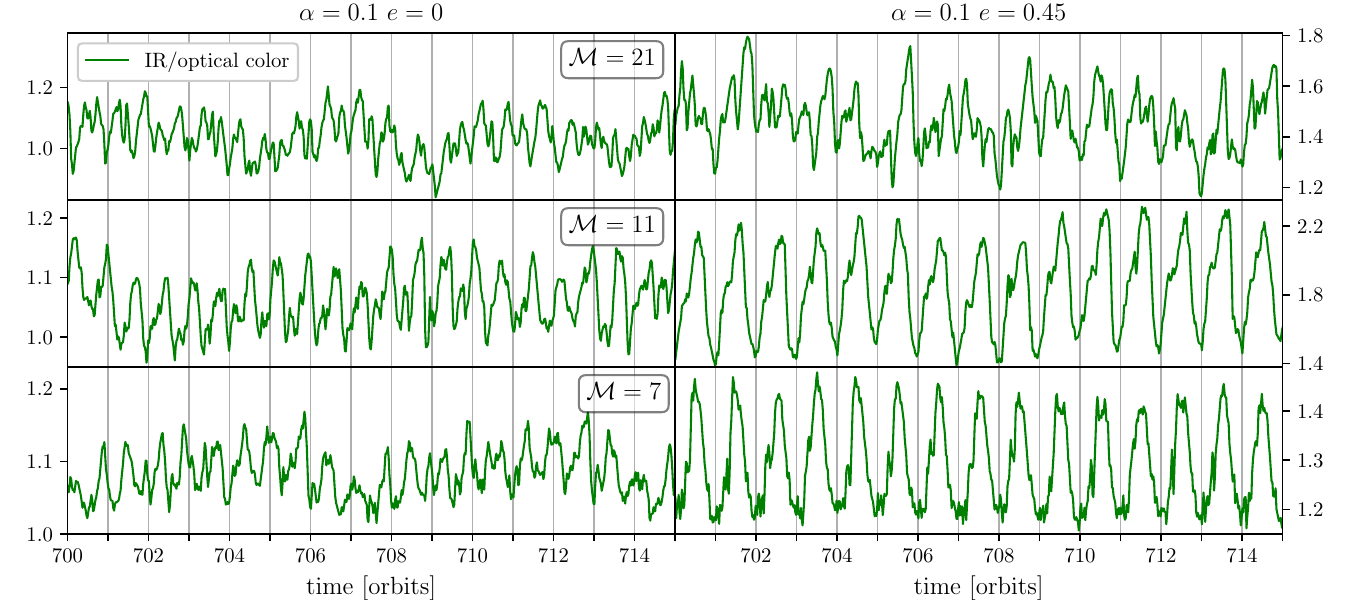}
\caption{Infrared-to-optical color as a function of time for circular ($e=0$, left column) and eccentric ($e=0.45$, right column) binaries. From top to bottom, rows correspond to Mach numbers 21, 11, 7. Each light curve includes corrections from our estimates of sub-sink and super-domain emission.} \label{fig:colorMach}
\end{figure*}

\begin{figure} 
\centering
\includegraphics[width=0.47\textwidth]{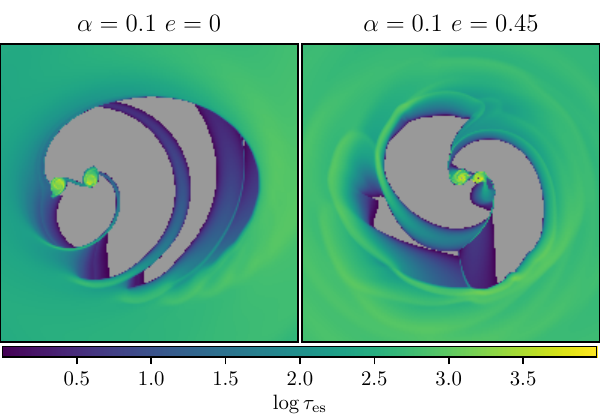}
\caption{Snapshots of the optical depth to electron scattering $\tau_{\rm es}$, displayed on log scale. Only the $\mathcal{M}=11$ case is shown, from our highest resolution runs. Left and right are circular and eccentric binaries, respectively. Regions with $\tau_{\rm es}<1$ are in gray.} \label{fig:taumap}
\end{figure}

In this section, we provide descriptions of some of our figures and make basic observations. We go into greater depth about the astrophysical implications of our results in \sref{sec:discuss}. When applicable, we quote ranges for our results, which accommodate all of our sensitivity tests presented in Appendix \sref{app:sens}. This is a conservative approach that is preferable to quoting simple averages of our test results, because the tests are not all equally important.

We have removed long-term ($\gtrsim\!20$-orbit) trends and applied sub-sink corrections to all light curves, and applied super-domain corrections to the infrared light curves, as described in \sref{sec:sub} \& \sref{sec:sup}. The super-domain correction to the infrared light curves amounts to roughly $2\times10^{42}\,$erg/s when the domain radius is $D=15 a$, and roughly $1.5\times10^{42}\,$ erg/s when $D=20 a$. The super-domain correction is uniform in time, whereas the sub-sink correction depends on the instantaneous accretion rate by each sink. For circular binaries, the proportion of time-varying luminosity that is resolved on the grid at a resolution of $\Delta x =0.01a$ and sink radius $r_{\rm sink}=4\Delta x$ is approximately 87\% for optical and 97\% for infrared. In our sink radius sensitivity tests, where $\Delta x = 0.005a$ and $r_{\rm sink}=0.04\Delta x$, the resolved portions improve to approximately 94\% for optical and 99\% for infrared. Reducing the sink further, so that $\Delta x = 0.0025a$ and $r_{\rm sink}=4\Delta x$, we resolve approximately 98.5\% of optical and 99.7\% of infrared. For eccentric runs with $e=0.45$, the trends for the resolved portions of time-varying luminosity as the sink radius is reduced is approximately $\lbrace 80\%, 91\%, 97\% \rbrace$ for optical and $\lbrace 97\%, 99\%, 99.7\% \rbrace$ for infrared. We therefore believe we are able to capture enough optical and infrared emission to make conclusions that are useful to observing campaigns.

To orient the reader, in \fref{fig:snaps} we display snapshots from our runs at different Mach numbers. All binaries orbit counter-clockwise in the figure. The runs with the highest Mach number ($\mathcal{M}(3a)=21$, top row) have notably disordered accretion streams and cavity walls, relative to the lower-Mach runs. Mach number is known to affect the stability of compressible laminar flows in ways which are difficult to anticipate in general. However, in a disk with constant-$\alpha$ viscosity, $\nu \propto \mathcal{M}^{-2}$ and thus higher-Mach flows in circumbinary disks are less stable due to the Reynolds number scaling with Mach number as ${\rm Re} \propto \mathcal{M}^2$. The eccentric runs (right column) are all shown at pericenter passage. White circles are drawn around each black hole corresponding to the Eggleton estimate of the Roche lobe radius \citep[specialized to $q=1$, see][]{eggleton1983}: $r_{\rm Egg} \equiv a \times 0.49/(0.6+\ln(2))$ for the circular binary, and we extend this to the eccentric case via the same adjustment that the pericenter distance would receive, namely $r_{\rm Egg} \rightarrow (1-e)r_{\rm Egg}$. The $\mathcal{M}\in\lbrace7,21\rbrace$ snapshots in \fref{fig:snaps} are from runs with resolution $\Delta x = 0.01 a$, whereas the snapshots from the $\mathcal{M}=11$ runs (middle row) are from our highest resolution runs with $\Delta x = 0.0025 a$. All snapshots shown have $r_{\rm sink}=4 \Delta x$.

\fref{fig:LCMach} displays optical and infrared light curves (in the system's cosmological rest frame) over a 15-orbit window for our fiducial viscosity $\alpha=0.1$, eccentricities $e\in \lbrace 0, 0.45\rbrace$, resolution $\Delta x = 0.01 a$, and sink radius $r_{\rm sink}=4 \Delta x$. From top to bottom, rows represent different initial Mach number profiles parametrized by $\mathcal{M}(3a)\simeq 21$, $\mathcal{M}(1.5a)\simeq 11$, and $\mathcal{M}(a)\simeq 7$; each case is a different representation of our target system accreting at $10\times$ the Eddington rate, as described in greater detail in \sref{sec:accrate}. We observe that the qualitative appearance of the light curves can vary substantially with Mach number, in particular the eccentric case with $\mathcal{M}=21$. In all cases, the light curves have a spiky appearance, and the visual impression of periodicity tends to decrease at larger Mach number. Also evident in \fref{fig:LCMach} is the presence of the ``lump'' period in the light curves from the circular binary. The lump has been seen in past studies, and has been described as an $m=1$ over-density that moves along the eccentric cavity wall surrounding the binary, with a period equal to several binary orbital periods. In addition to the lump periodicity, we observe in \fref{fig:LCMach} a faster modulation on the order of the orbital period. The orbital modulation is especially obvious in the eccentric case, for $\mathcal{M}\in\lbrace 11, 7\rbrace$. We also observe that variability in the light curves is substantial (which is purely hydrodynamic, in contrast with a Doppler variability). For the circular binary, the root-mean-squared (RMS) variability is $3.3-8.5\%$ in the optical band, and $0.90-2.6\%$ in the infrared. For the eccentric binary, it is larger: $7.7-15\%$ in the optical and $1.9-3.5\%$ in the infrared (see Tables~\ref{tab:circ_var} \& \ref{tab:ecc_var}). Note that the peak-to-trough difference is roughly $3\times$ the RMS variability. We compare this variability to Doppler brightening in \sref{sec:dop}.

\fref{fig:specMach} provides quantitative corroboration of our qualitative judgements about \fref{fig:LCMach}. The panel organization is the same, except Lomb-Scargle periodograms (normalized to a maximum value of 1) are plotted versus frequency in units of orbits$^{-1}$. The lump frequency is obvious for the circular binary, having a range of values corresponding to a period of $\simeq 5-10$ binary orbital periods. We also see prominent peaks near the orbital frequency of 1, although for the circular binary the peak is quite obviously at $\simeq 1.4$ orbits$^{-1}$. The peak at $\simeq 1.4$ orbits$^{-1}$ decreases towards the orbital frequency in our sink-shrinking tests. This behavior is shown in \fref{fig:ftconverge}. But to be conservative about our conclusions, the range of frequencies for the near-orbital modulation for the circular binary is quoted as $1.0-1.5$ in Appendix \sref{app:sens}, and we refer to this modulation as the ``fast'' frequency $f_{\rm fast}$ or the ``near-orbital'' frequency. A similar frequency of $\unit[1.46]{orbits^{-1}}$ was observed in past studies \citep[e.g.][]{Roedig+2012, noble+2012, shi+2015}, and was interpreted as a beat frequency (for example, $2(f_{\rm bin} - f_{\rm lump})$, where $f_{\rm bin}$ and $f_{\rm lump}$ are the binary and lump frequencies). Given the lump frequencies we find, and the fact that the near-orbital frequency varies in our sink-shrinking tests, our $f_{\rm fast}$ does not appear to be such a beat frequency. Instead, our sink-shrinking test shows that our $f_{\rm fast}$ is a phenomenon dependent on gravitational softening, sink size, sink rate, and/or resolution (since our sink-shrinking test varies all of these parameters simultaneously). We found in our investigations that $f_{\rm fast} - f_{\rm orb}$ is the precession frequency of the minidisks, primarily driven by gravitational softening. Although gravitational softening is required for stability in Newtonian simulations, in the 2-dimensional thin disk setting, gravitational softening is required to account for the vertically-integrated, plane-parallel component of the gravitational force on the disk. Thus, if a result depends on softening, it does not necessarily follow that the result is an artifact. Instead, the result may depend on the disk's vertical structure. We will report on this precession phenomenon and its dependence on softening in greater detail in future work.

The periodograms for the eccentric binary have a comparative absence of the lump frequency. Interestingly, however, it is not completely absent, especially in the infrared band. The $\mathcal{M}=21$ case has several peaks at low frequency, which may indicate a lump-like phenomenon. There is a qualitative suggestion of the presence of a lump in the corresponding panel of \fref{fig:snaps}. Since higher Mach number flows are inherently less stable, it is arguably not surprising that a lump could appear at higher Mach numbers around the eccentric binary. Further investigation of lump-like periodicity in high-Mach number disks around eccentric binaries is beyond our present scope, but it suggests that some eccentric binaries can be circular binary ``imposters.'' In the eccentric $\mathcal{M}\in\lbrace 11, 7\rbrace$ cases, the first harmonic of the orbital frequency appears clearly; harmonics often arise as the Fourier-representation of periodic pulses that do not have a purely sinusoidal shape (as is the case in the corresponding panels of \fref{fig:LCMach}).

To gain insight into where modulated emission comes from, we present \fref{fig:ftMD} for the $\mathcal{M}=11$ case only, showing normalized Lomb-Scargle periodograms for the infrared emission coming from the minidisks (regions within $r_{\rm peri}/2$ from each black hole) and emission coming from elsewhere (mostly the accretion streams and cavity wall). We refer to the latter emission as the ``cirumbinary'' (CBD) emission. The optical case (not shown) is similar, so we omit it in order not to clutter \fref{fig:ftMD}. Other Mach numbers are also similar, so we believe \fref{fig:ftMD} is representative. For the circular binary, we observe that the net minidisk emission is dominated by the near-orbital period, whereas the accretion streams and/or cavity wall are dominated by the lump periodicity. In the eccentric case, all emission is dominated by the orbital period. Spatial maps of the optical and infrared luminosity are displayed in \fref{fig:lummaps}, showing bright minidisks, a diffuse glow in the lump, and narrow bright features along accretion streams and the shock-heated cavity wall. For the circular binary, just as the minidisk emission is not modulated significantly on the lump period, nor are the accretion rates as registered by the sinks. This means that lump periodicity does not transmit to the jet emission via $\dot{M}$. However, it is conceivable that lump periodicity could manifest in the jet via up-scattering of photons emitted from the CBD.

We also note that for the circular binary, a beat frequency appears in the emission from the individual minidisks (not shown). This beat frequency is $f_{\rm fast} - f_{\rm bin}$, and it is completely out-of-phase between the minidisks, which explains why it does not appear in the periodogram of the net minidisk emission in \fref{fig:ftMD}. It has a value similar to a lump frequency ($\simeq \unit[0.16]{orbits^{-1}}$ in the top panel of \fref{fig:ftMD}), but is clearly distinct from the lump frequency $f_{\rm lump} \simeq \unit[0.11]{orbits^{-1}}$. Since $f_{\rm fast}$  moves towards $f_{\rm bin}$ in our sink-shrinking test, the beat frequency $f_{\rm fast} - f_{\rm bin}$ seems to approach zero. Since our sink-shrinking test varies the resolution, sink size, sink rate, and gravitational softening length simultaneously, the beat frequency must be a function of this subset of parameters. We will also investigate this in future work.

In \fref{fig:colorMach} we plot the infrared-to-optical ratio (``color''), with the same panel organization as \fref{fig:LCMach} \& \fref{fig:specMach}. The lump periodicity for the circular binary is noticeably suppressed in comparison with \fref{fig:LCMach}. To the extent that the lump periodicity is eliminated in the color, that indicates that the lump signal is in-phase between the optical and infrared bands, and with a similar magnitude (relative to the average) and shape in both bands. We describe this as ``achromaticity'' of the lump periodicity. In Table~\ref{tab:circ_color}, we quantify the degree to which the prominence of lump periodicity (in relation to the near-orbital frequency) decreases when going from the individual infrared and optical bands to the color: the prominence of lump periodicity decreases by $46-83\%$ compared to optical and $80-99\%$ compared to infrared. 
This is interestingly different from the well-established ``bluer when brighter'' chromaticity of general stochastic AGN variability.

For the eccentric binary, for $\mathcal{M}\in\lbrace 11,7\}$, although the orbital modulation in different bands is clearly in-phase (peaking at pericenter, i.e.~integer values of orbits, see \fref{fig:LCMach}), the magnitude and shape of the two bands is nonetheless different enough that the color peaks mid-orbit (at apocenter, i.e.~at times halfway between integer orbits). The average color is useful to compare to the emission from a steady circumsingle disk around a black hole of mass equal to the binary's mass. The circumsingle disk color is 0.94, whereas a circular binary has a modest enhancement to $0.97-1.2$, and an eccentric binary has an even greater enhancement to $1.3-2.1$. The color variability on the near-orbital time scale is also a notable observable: RMS of $2.4-6.2\%$ for circular binaries, $6.3-12\%$ for eccentric binaries. See Tables~\ref{tab:circ_color} \&~\ref{tab:ecc_color}. A direct comparison to the complex phenomenology of AGN observations (see e.g. \citealt{padovani+2017}) is involved, and is beyond the scope of the present study.

\begin{figure*} 
\centering
\includegraphics[width=1\textwidth]{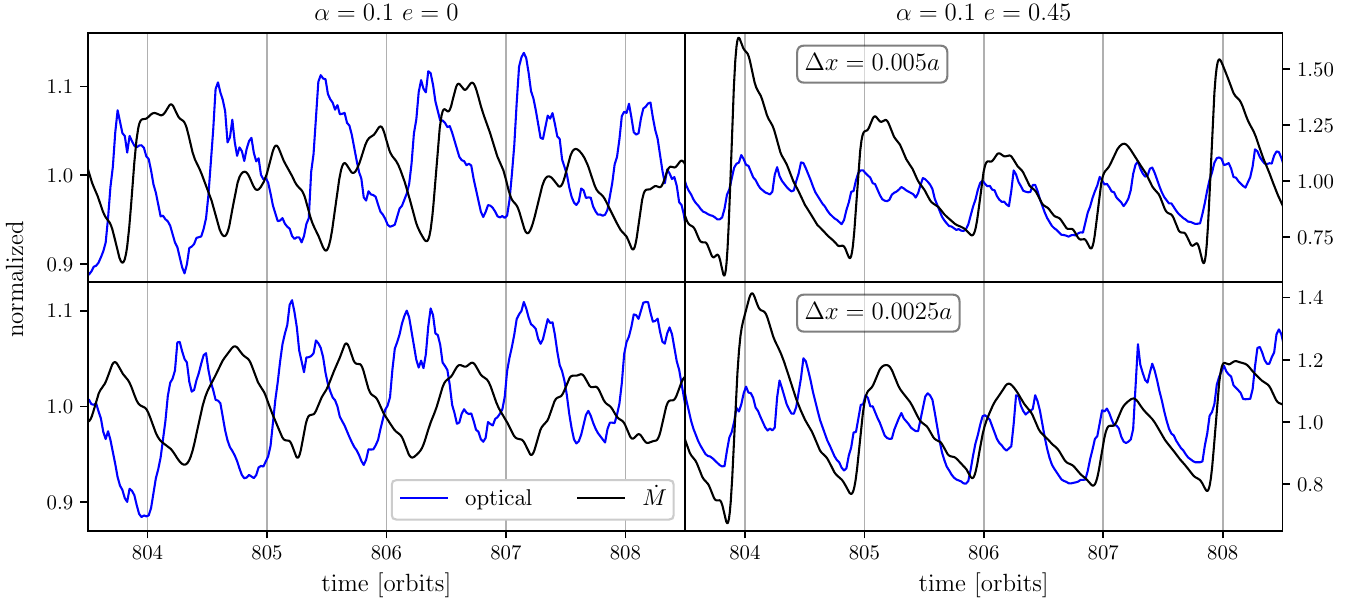}
\caption{Optical LC and total $\dot{M}$ normalized by their averages for the $\mathcal{M}=11$ high-resolution sink-shrinking study (test label: $r_{\rm sink}$ in Appendix \sref{app:sens}). Top row: $\Delta x = 0.005 a$, bottom row: $\Delta x = 0.0025 a$. Both cases have $r_{\rm sink}=4 \Delta x$. Left column: circular binary, right column: eccentric binary. Significant lags are apparent in the circular case, but absent in the eccentric case. The lags appear consistent between the two resolutions, suggesting the lags are well-resolved.} \label{fig:Mdotlag}
\end{figure*}

In \fref{fig:Mdotlag} we plot the optical LC and total accretion rate $\dot{M}$ from our sink-shrinking study (test label: $r_{\rm sink}$ in Appendix \sref{app:sens}). The accretion rates are measured by the sink terms, in a neighborhood of the black holes. The rows show resolutions $\Delta x = 0.005 a$ (top) and $\Delta x = 0.0025 a$ (bottom). The lag between the optical LC and $\dot{M}$ appears converged, and is highly in-phase for the eccentric binary (right column); for the circular binary (left column), they are nearly out-of-phase. The infrared LC (not shown) is in-phase with the optical LC. If jet emission is well-predicted by $\dot{M}$, then this is a tantalizing hint of a smoking-gun signature of eccentricity, in SMBHB candidates with well-differentiated jet and disk emission. Without having performed sink-shrinking tests for all of our sensitivity tests in Appendix \sref{app:sens}, this result is not up to the same standard of evidence as our other results tabulated in Appendix \sref{app:sens}. However, it stands to reason that the smaller minidisks in the eccentric case result in less buffering of incoming perturbations, thereby allowing a greater degree of synchronicity between the consequent minidisk heating (and thus emission) and accretion by the black holes.

Lastly, in order to assess our assumption that the gas is optically thick everywhere, we display the electron scattering optical depth $\tau_{\rm es}$ of the gas in \fref{fig:taumap}, where regions with $\tau_{\rm es}<1$ have been greyed out. The electron scattering optical depth $\tau_{\rm es}$ is a lower bound for the effective optical depth (see \sref{sec:disk}). To compute this, as per \sref{sec:accrate}, we first scaled the surface density down to the general level of our target model via the approximate map
\begin{eqnarray}
    \Sigma &\rightarrow & \Sigma \times \left(\frac{10}{A}\right)
\end{eqnarray}
This map follows from these relations:
\begin{eqnarray}
    \dot{M} &\rightarrow& \dot{M}\times\left(\frac{10}{A}\right)\\
    \dot{M} &=& 3\pi\Sigma\nu \\
    \nu &\rightarrow& \simeq \nu,
\end{eqnarray}
and ansatz (for some $n$):
\begin{eqnarray}
    \Sigma &\rightarrow& \Sigma \times \left(\frac{10}{A}\right)^n.
\end{eqnarray}
The only regions with $\tau_{\rm es}<1$ are in the low-density cavity, from which a subdominant amount of luminosity is expected.

\section{Discussion} \label{sec:discuss}

In this section, we discuss our results in greater depth, organizing into two categories: signatures of a binary in general, and signatures of an eccentric binary in particular. We then discuss the effect of Doppler brightening, which is often a central consideration when modeling binary quasar light curves \citep[see e.g.][]{dorazio1+2015, Charisi2021}. Finally, we discuss, as an illustrative example, a particular quasar with a claimed quasi-periodic light curve, to which our results may be relevant in the future. The optical and infrared signatures we discuss presuppose that the emission from the system is either dominated by the thermal disk emission \citep[rather than the jet(s), a dusty torus, etc.~-- see][]{padovani+2017}, or otherwise that the thermal disk component can be separated out through spectral modeling.

Throughout this section, we refer to root-mean-square (RMS) variabilities. RMS variability is a readily recognized measure, but keep in mind that the peak-to-trough variability (which is more easily judged visually in plots) is roughly $3\times$ the RMS variability.

\subsection{Signatures of a binary} \label{sec:binevi}
One signature of binarity that would be very conspicuous with a sufficiently long temporal baseline is the simultaneous presence of two significant periodicities separated by a factor of $\simeq 4.5-16$. These would be the orbital (or near-orbital) periodicity and the lump periodicity. Harmonics of either of these periods are likely, but harmonics are such a generic phenomenon (encoding non-sinusoidal pulse shapes) that they are not strong evidence of binarity by themselves. Even more conspicuous would be different relative weights of the lump versus near-orbital periodicity in different bands. We tend to find that the lower energy bands (infrared) tend to have an over-representation of the lump periodicity, and an under-representation of the near-orbital periodicity, relative to higher energy bands (optical).

We generally find that optical variability on the orbital time scale is significantly larger than infrared; we found that infrared RMS variability is $\simeq 0.90-3.5\%$, whereas optical RMS variability is $\simeq 3.3-15\%$ (see Tables~\ref{tab:circ_var} \&~\ref{tab:ecc_var}). This itself is a signature of binarity, but may not be sufficiently specific to be convincing.  Note that stochastic AGN variability amplitudes are also generally larger in bluer bands, but the ratio between optical and infrared is significantly below the factor of $3-4$ we found for binaries here. If ordinary quasar light curve rms variability is on the order of 10\%, then the periodic hydrodynamic variability due to a binary may very well be obscured, especially without a temporal baseline extending for many binary orbital periods.

The preponderance of our simulations show lags between optical and infrared that are consistent with zero (but see Tables~\ref{tab:circ_lag} \&~\ref{tab:ecc_lag} for a few exceptions). Our temporal cadence is roughly 2\% of an orbit, which translates into lags $\lesssim \unit[1]{week}$ compared to our orbital period of 1 year (in the source frame). We measure lags by computing the discrete correlation function~\citep[][]{Edelson+1988} between the two signals, and seeing at which lag the first local maximum occurs; correlations are quoted as the value at that local maximum. We find the level of correlation between optical and infrared to be $\simeq 0.64-0.96$. See Tables~\ref{tab:circ_lag} \&~\ref{tab:ecc_lag}. If ordinary quasar variability is of order 10\% and uncorrelated, one would reasonably expect a decrease in these correlations by an amount on the order of 0.1. Even so, the level of correlation between different bands for circumbinary emission is quite high.

The infrared-to-optical ratio (``color'') is on average $\simeq 0.97-2.1$, above the corresponding circumsingle disk color of $\simeq 0.94$ (see Tables~\ref{tab:circ_color} \&~\ref{tab:ecc_color}). This may be an important signature of binarity, but it would be important to understand how much other processes in circumsingle disks (e.g.~Lightman-Eardley or ionization instabilities) could enhance the color above $\simeq 0.94$. Chromatic variability from a binary can be quite large (we find $\simeq 2.4-12\%$), and periodicity is generally present. For circular binaries, the prominence of lump periodicity compared to the near-orbital frequency decreases substantially when going from specific bands to the color. For example, the lump prominence decreases by $80-99\%$ from the infrared band, and by $46-83\%$ from the optical band. See Table~\ref{tab:circ_color}. This degree of achromaticity is unusual for AGN. It is worth mentioning a very recent work \citep{Negi+2021} finding chromatic variability in blazars, in particular. Although our results apply to more general AGN, we discuss applications of our results to blazars in \sref{sec:sources}.

\subsection{Signatures of an eccentric binary} \label{sec:eccbinevi}
Most of our simulations show a large relative lack of lump periodicity in eccentric binaries versus circular binaries. This may signal eccentricity, but some of our simulations show that it is possible for eccentric binaries to be circular imposters by exhibiting significant lump periodicity. Thus, relative strengths of lump versus near-orbital periodicity in different bands are tentative ways of distinguishing eccentric and circular binaries which require further elucidation.

In optical, infrared, and in color, eccentric binaries have systematically higher variability on the orbital time scale. This is especially true of optical, with circular binaries exhibiting $\simeq 3.3-8.5\%$ variability and eccentric ones exhibiting $\simeq 7.7-15\%$ variability. See Tables~\ref{tab:circ_var},~\ref{tab:ecc_var},~\ref{tab:circ_color}, and~\ref{tab:ecc_color}. The average color is also significantly larger for eccentric binaries: $\simeq 1.3-2.1$ for eccentric versus $\simeq 0.97-1.2$ for circular. On the other hand, we do not find a significant difference in correlations between infrared and optical emission.

As discussed at the end of \sref{sec:results}, \fref{fig:Mdotlag} shows a tantalizing result that the jet emission may have a significant lag with respect to low-energy disk emission for circular binaries, but not for eccentric binaries. Deeper investigations are warranted, i.e.~performing sink-shrinking tests for all of our sensitivity tests in Appendix \sref{app:sens}. Such a suite of simulations would be significantly more costly than those we have already performed, and we intend to do so in future work.

\subsection{Doppler brightening} \label{sec:dop}

If the binary is not viewed face-on, then emitting gas parcels will appear Doppler brightened (or dimmed). In particular, the net motion of gas in each minidisk will cause Doppler brightening tied to the orbital motion of the black holes. Doppler brightening is a relativistic effect appearing first at order $v_\parallel/c$, where $v_\parallel$ is the net line-of-sight velocity of the minidisk. Roughly speaking, the binary orbital motion only brightens the minidisks, since they have net motion with their respective black holes. Other emitting gas, such as the cavity wall, is not moving as quickly. In our simulations, we consider minidisk emission to be coming from within a distance $r_{\rm peri}/2$ from its black hole -- such regions are drawn in \fref{fig:snaps}.

For our particular binary parameters, the orbital velocity is $\simeq 0.02c$ for the circular binary, and $\simeq 0.033c$ for the $e=0.45$ binary at pericenter. We have checked how large the Doppler brightening effect is for our system, and found it to be negligible ($\lesssim 1$\% effect on light curves for edge-on observers). However, it is of interest to estimate how large the orbital velocities would have to be for Doppler brightening to be important. We do so in this section.

Brightness variability coming from the Doppler effect competes with hydrodynamic effects. We seek to estimate how large $v/c$ must be for hydrodynamic and Doppler variability to have equal amplitude (see also~\citealt{dorazio1+2015}). To this end, we schematically decompose the total observed luminosity as
\begin{equation}
    L_{\rm tot}^{\rm obs}(t) = \langle L_{\rm tot} \rangle \left( 1 + \delta_{\rm Doppler}(t) + \delta_{\rm hydro}(t) \right), \label{eq:lumschem}
\end{equation}
where $\langle L_{\rm tot} \rangle$ is the average intrinsic total luminosity, and the relative variabilities $\delta_{\rm Doppler}(t)$ and $\delta_{\rm hydro}(t)$ are caused by Doppler and hydrodynamic effects, respectively. Our aim is to compare the amplitudes of these two terms. Two considerations must be made. First, only the fraction of the total luminosity produced by the minidisks is subject to Doppler brightening from binary orbital motion. Second, the net change in observed flux includes Doppler-brightening of the approaching minidisk, in addition to Doppler dimming of the receding one. These effects tend to oppose each other, and would cancel at first order in $v/c$ if the minidisks both had the same power-law spectra. Let $b \equiv \langle L_{\rm MD} \rangle / \langle L_{\rm tot} \rangle$ quantify the fraction of average emission that originates in the minidisks (and is thus susceptible to Doppler modulation), and let $f\equiv \langle L_2 \rangle / \langle L_1 \rangle$ quantify the average minidisk brightness asymmetry (defined such that $f\leq1$), where $\langle L_1 \rangle$ and $\langle L_2\rangle$ are average luminosities for the individual minidisks. We relegate a detailed calculation to Appendix \sref{app:dop}, and provide the main results of the calculation here.

To maximize the Doppler variability, we consider edge-on observers, resulting in
\begin{eqnarray}
    \delta_{\rm Doppler}(t) \simeq 3.44 \,b \left(\frac{1 - f}{1 + f}\right) \frac{v}{c} \sin(\Omega_{\rm bin} t),
\end{eqnarray}
where $v$ is the orbital velocity. The factor of $3.44$ comes from an estimate of the power-law spectral index (see Appendix \sref{app:dop}). The hydrodynamic variability has a minidisk contribution $\delta_{\rm MD}(t)$, and a contribution from elsewhere -- we call the latter the ``circumbinary'' (CBD) contribution $\delta_{\rm CBD}(t)$, so that $\delta_{\rm hydro}(t) \equiv \delta_{\rm MD}(t) + \delta_{\rm CBD}(t)$. These contributions are estimated as
\begin{eqnarray}
    \delta_{\rm MD}(t) &=& 1.5 \, b \, A_{\rm MD}(t) \\
    \delta_{\rm CBD}(t) &=& 1.5 \, (1 - b) \, A_{\rm CBD}(t),
\end{eqnarray}
where $A_{\rm MD}(t)$ and $A_{\rm CBD}(t)$ are quasi-periodic modulations with amplitudes ($\equiv \bar{A}_{\rm MD}$ and $\bar{A}_{\rm CBD}$) equal to the root-mean-square (RMS) variability of unboosted minidisk and CBD emission, respectively. These two contributions tend to be coherent with each other, but to be conservative about the amplitude of $\delta_{\rm hydro}$, we take it to be the maximum of the amplitudes of $\delta_{\rm MD}$ and $\delta_{\rm CBD}$. The amplitude of $\delta_{\rm Doppler}$ then equals $\delta_{\rm hydro}$ when
\begin{eqnarray}
    \!\!\!\!v/c \simeq \frac{1.5}{3.44\, b} \left( \frac{1+f}{1-f} \right)\max\lbrace b\bar{A}_{\rm MD}, (1-b) \bar{A}_{\rm CBD} \rbrace. \label{eq:dopvmin}
\end{eqnarray}
We compute this estimate of $v/c$ for all of our simulations in Appendix \sref{app:sens}. By taking the minimum $v/c$ across all of our simulations,
\begin{itemize}
    \item $e=0$, optical: $v/c \gtrsim 0.16$
    \item $e=0$, infrared: $v/c \gtrsim 0.14$
    \item $e\gtrsim 0.45$, optical: $v/c \gtrsim 0.12$
    \item $e\gtrsim 0.45$, infrared: $v/c \gtrsim 0.16$.
\end{itemize}
All of these line-of-sight velocities are rather high; by comparison, the orbital velocity of our circular binary is $v/c \simeq 0.02$, or $v/c\simeq0.033$ at pericenter for our $e=0.45$ binary. Most binaries would be viewed at some intermediate angle, as opposed to edge-on, which increases the required orbital velocity to achieve parity between Doppler brightening and hydrodynamic variability by a few tens of percent.

Our results indicate that Doppler modulation of brightness is generally sub-dominant to hydrodynamic variability in accreting equal-mass binaries with disk Mach numbers $\mathcal{M}\lesssim 20$, even when the binary has significant eccentricity. These results are consistent with findings from \cite{Tang2018}, which were limited to circular binaries. On the other hand, we note that lower-mass ratio binaries have smaller hydrodynamical variability, particularly for $q\lesssim 0.05$~\citep{Farris+2014,Dorazio+2016}, and Doppler variability can be dominant for these~\citep{dorazio1+2015}. Exploring how the relative importance of Doppler modulation and hydrodynamic variability translates to unequal-mass systems, and to higher Mach numbers, will be the subject of future work.

If periodic Doppler brightening signals are most detectable for low mass ratio binaries or higher Mach number disks, then their detection would serve to constrain those parameters. Excitingly, if mass ratio can be constrained by other means (e.g.~GWs), then periodic Doppler brightening signals might provide a rare constraint on disk Mach numbers, which in turn constrains combinations of other disk parameters like accretion rate, turbulent viscosity, and surface density.

\subsection{Application to periodic blazars} \label{sec:sources}
Since we are primarily focusing on the resolved infrared and optical thermal emission from the disk, our predictions are most relevant for supermassive black hole binaries in AGN for which the jet emission components are either subdominant or can be modeled out. For this reason, predictions about thermal disk emission are not directly relevant to BL Lacertae objects, since they are more dominated by jet emission at all wavelengths. However, we do predict properties of the jet emission to the extent that they track accretion rates. Energy considerations motivate a schematic relationship between jet power $L_{\rm jet}$ and the accretion rate,
\begin{eqnarray}
    L_{\rm jet} \approx \eta \dot{M} c^2, \label{eq:jetvsmdot}
\end{eqnarray}
where $\eta$ is the efficiency. This relationship is generally not expected to hold instantaneously, but rather on average \citep[e.g.][]{paschalidis+2021, combi+2021}. However, for the binaries considered in this work, the periodic variability of $\dot{M}$ takes place on the binary orbital time scale of 1 year, whereas the characteristic orbital time near the black holes is a few minutes. Thus, given how slow the accretion rate modulations are for our binaries, it is sensible to expect Eq.~\eqref{eq:jetvsmdot} to hold in a quasi-stationary sense in our case.

In this section, as a basis for an illustrative discussion, and to limit our scope, we focus primarily on a specific flat spectrum radio quasar (FSRQ) for which there is a recent claim of quasi-periodic oscillations at a period of $\simeq 2.6$ years in $\gamma$-rays: PKS 0208-512~\citep{penil+2020}. Only a few cycles are present in the data, therefore stochastic flaring is not precluded. However, it serves as a useful case to explore ideas on how our results could be applied to binary blazars. This object is at a redshift of $z\simeq1$ \citep{0208redshift}, which means the source frame periodicity is $\simeq 1.3$ years. If this periodicity corresponds to the orbital period of a putative binary, then it is similar to the fiducial model we studied in this work. The total black hole mass has been estimated based on various assumptions to be $(6-25) \times 10^7 M_\odot$ \citep{stacy0208mass, yang0208mass}, which is in the range of $\simeq8-30$ times the total binary mass we studied in this work. The disk effective temperature of PKS 0208-512 can be made similar to the system we studied in this work via a corresponding scaling down of the accretion rate, to $\simeq (0.01 - 0.15) \times \dot{M}_{\rm Edd}$. This would make many of our predictions applicable (see \sref{sec:caveats} for more discussion of this point). However, one still has to isolate the thermal disk emission, which could be done through model fits of the observed broadband spectrum. In addition to $\gamma$-ray data, there exists X-ray, optical, and infrared data for PKS 0208-512, and they have been analyzed in a number of contexts \citep[see e.g.][]{chatterjee+2013, khatoon+2021}. It has been suggested that the disk emission is observable when the jet activity is low \citep{chatterjee+2013}. However, we are unaware of a periodicity analysis in bands other than $\gamma$-rays \citep{penil+2020}.

Using Eq.~\eqref{eq:agw}, we can check that the above putative binary with total mass $\simeq (6-25) \times 10^7 M_\odot$ is in the GW-driven regime for a wide range of parameter values. One can therefore expect that it is circularizing. If it has circularized sufficiently, then our predictions for equal-mass circular binaries may apply. Assuming PKS 0208-512 hosts a near-equal mass, near-circular binary, we predict that the thermal disk emission in infrared and optical (or far-infrared and near-infrared in the observer frame due to cosmological redshift $\nu \rightarrow \nu/(1+z)$) should exhibit in-phase ($< 20$ days in the observer frame) brightness modulations on near-orbital time scales, with a prominent lump period of $\simeq 4.5 - 16 $ times the orbital period. The lump period will be more prominent (relative to the orbital period) in lower-energy disk emission. Given the semi-major axis of $a\simeq \unit[(2-5)\times 10^{-3}]{pc}$ implied by the inferred binary mass range and a 1.3-year orbital period, the maximum orbital velocity is on the order of $v/c\simeq 0.03-0.08$ for a circular binary and $v/c\simeq 0.05-0.13$ for an eccentricity of $e=0.45$. Since PKS 0208-512 is a blazar (i.e.~viewed nearly face-on), the line-of-sight orbital velocities are likely well below the levels in \sref{sec:dop}. Thus, Doppler brightening on the orbital time scale is very likely irrelevant, so we can predict that a binary would produce spiky orbital periodicity, with the near-infrared light curves (observer frame) having more significant ($3.3-8.5\%$) RMS variability in comparison to far-infrared ($0.90-2.6\%$).

Keeping our caveats from \sref{sec:caveats} in mind, and the assumptions above, we predict that periodic modulations in the jet emission will be significantly out-of-phase with disk emission. Absence of the lump frequency in our measurements of the $\dot M$ time series also implies that if modulations of $\gamma$-ray brightness reflect the intrinsic jet power \citep[as per e.g.][]{combi+2021}, the jet emission should not be modulated by the lump. On the other hand, the lump periodicity has been observed to appear in $\dot{M}$ in past work using the same non-isothermal equation of state as ours \citep[e.g.][]{farris+2015b}. It would be interesting to understand the conditions under which lump periodicity does and does not transmit to accretion rates, including the effect of the sink size, since that informs whether one should expect jet power to be systematically modulated on that time scale.

On the other hand, if modulations of $\gamma$-ray brightness instead reflect the supply of seed photons to the inverse Compton process in the jet, then some long-term trends in blazar $\gamma$-ray light curves might reflect lump periodicity in that external supply of seed photons. For example, it has recently been noted that PKS 0208-512 has had an increase in $\gamma$-ray activity beginning around 2019 \citep[see e.g.][]{khatoon+2021}, and the trend in flare amplitudes suggests a lump-like (e.g.~$\simeq5$-orbit) period. So it is worth considering jet brightness modulations arising from the supply of seed photons from the disk. In this scenario, one would expect the inverse Compton component of the blazar spectral energy distribution to have an imprint of lump periodicity (whereas the lower energy sychrotron component would have such an imprint only if the lump periodicity transmits to the accretion rates).

In this external inverse Compton scenario, since the cavity is the region whose emission is primarily modulated at the lump period (see \fref{fig:ftMD}), the time delay between CBD brightening and upscattering of the resultant photons by the jet would be at least on the order of the light-travel time from the cavity wall to the black holes. This delay is of order $\simeq 1\%$ of an orbital period for the putative binary in PKS 0208-512.

If the lump period is indeed being transmitted to jet emission, then in addition to predicting flares in early/mid-2021, early-2024, and mid-2026, the amplitude of those flares will show evidence of reverting to the low-activity state similar to the years $2009-2017$. 

An ``orphan'' flare from PKS 0208-512 in the optical-near-infrared band (i.e.~no $\gamma$-ray counterpart) was reported between the two $\gamma$-ray flares that bracket the interval from $2008-2011$ \citep{chatterjee+2013}. Orphan flares in low-energy bands could be explained by a binary, since one would expect flares in disk emission that are out-of-phase with the jet (either due to the binary being circular, or even for eccentric binaries since the pulse substructure of disk emission can be distinct from $\dot{M}$; see \fref{fig:Mdotlag}). Since the low-energy flares occur while the jet is not flaring, the disk emission would be more visible at that time than otherwise. If binarity is the cause of an orphan flare, we would predict that the orphan low-energy flares have a thermal spectrum (i.e.~originating in the disk), whereas the low-energy flares coincident with $\gamma$-rays would be non-thermal (i.e.~originating in a jet).

Lastly, the BL Lacertae object PKS 2155-304 has several reports of 1.7-year $\gamma$-ray periodicity, but different studies disagree about whether optical data has the same 1.7-year periodicity or roughly half of that (0.87 years) \citep[e.g][]{zhang+2014, sandrinelli+2014, sandrinelli+2016, covino+2020, penil+2020, bhatta2021}. If optical flares from the disk around a binary are out-of-phase with jet flares, one would expect a periodicity in the optical band at roughly twice the jet periodicity. Since the flares in disk emission may be weaker than the jet, they may be harder to discern in the data, which might explain the mixed claims of 0.87-year and 1.7-year periodicity in the literature.

\subsection{Caveats} \label{sec:caveats}
In this section, we discuss some caveats, focusing on the ones we believe are most important for interpreting EM signatures of SMBHB candidates.

Firstly, when generalizing our results about optical and infrared emission to other binary parameters, one must pay special attention to the overall effective temperature of the gas. If the effective temperature of our system were scaled down, then infrared and optical emission would come from deeper inside the gravitational wells of each black hole. Then, the lump periodicity would feature less prominently, and the fraction of emission coming from the minidisks would be increased. That effect would be reversed if the effective temperature were instead scaled up. Thus, one ought to think spatially, by associating our results about infrared and optical emission with whatever emission is coming $\simeq 7.6-28\%$ and $\simeq 37-84\%$ from the minidisks, respectively (see $b$ in Tables~\ref{tab:circ_MD} \&~\ref{tab:ecc_MD}); which band of emission that corresponds to will depend on the accretion rate $\dot{M}$, binary mass $M$, and orbital semi-major axis $a$ roughly via $T_{\rm eff}^4 \sim M \dot{M} a^{-3}$. However, if the amplitude of periodic signals depend on such parameters, then our results should not be taken as generic. One also must apply a cosmological redshift to the bands ($\nu \rightarrow \nu/(1+z)$), since our results are quoted in the cosmological rest frame of the source.

Secondly, we caution against over-interpreting the shapes of specific pulses seen in \fref{fig:LCMach}. The pulse shapes for a given model change from one cycle to the next, and also vary in character between models with different nominal Mach numbers. Our tests show reliable results about the frequency content of the pulse time series (i.e. near-orbital and lump frequencies having values within a certain range, see Table~\ref{tab:circ_freq}), and it is safe to say that the pulse shapes resulting from hydrodynamic processes are probably ``spiky,'' and not perfectly periodic. 

Furthermore, the Mach number of the gas depends strongly on the accretion rate. For lower accretion rates, the Mach number can easily become greater than 100; for example, if our target system were accreting at $0.1\times\dot{M}_{\rm Edd}$ instead of $10\times\dot{M}_{\rm Edd}$ (all else being equal), then the initial condition would have $\mathcal{M}(a)\simeq240$. Our \fref{fig:LCMach} suggests that the visual appearance of light curves depends very strongly on Mach number, becoming noisier at higher Mach. One wonders whether light curves from very high-Mach number circumbinary accretion disks would present any obvious periodicity at all, unless from Doppler modulation. However, even though the Mach numbers we have simulated are in a narrow range of possible values, the Mach numbers we have simulated are physically reasonable. In particular, the Mach numbers we simulated do not require super-Eddington accretion rates; the Mach profile of the $\alpha$-disk models we use are nearly invariant when simultaneously scaling the black hole mass up by a factor of $n$ and the accretion rate (expressed in multiples of Eddington) down by a factor of $n$. Thus, for example, a binary with $M=10^9\,M_\odot$ accreting at 0.1$\times$ the Eddington rate would have similar Mach number as the systems we study in this work.

Next, as we have mentioned above, we are primarily predicting the thermal disk emission.
Infrared wavelengths are expected to be heavily contaminated by dust near the quasar \citep[e.g.][]{padovani+2017}, which must be taken into consideration when applying our results to observations. However, since the specific bands we present should be thought of more generally as whatever emission is coming $\simeq 7.6-28\%$ and $\simeq 37-84\%$ from the minidisks, dust will not be a contaminant in general (in particular, for sufficiently higher binary mass and/or compaction).

We can only infer properties of the jet emission to the extent that it is predicted by $\dot{M}$ \citep[which has some theoretical support, e.g.][]{paschalidis+2021, combi+2021} or by the disk supply of seed photons. Our inferences about $\dot{M}$ are not up to the same standard as our other results quoted in Tables~\ref{tab:circ_var} through~\ref{tab:ecc_MD}, because that was not a primary design goal of our study. We believe that our conclusions involving $\dot{M}$ would be at the same standard as our infrared and optical conclusions if we performed sink-shrinking studies in all of our sensitivity tests in Appendix~\ref{app:sens}. Currently, we only performed a sink-shrinking study for our fiducial $\mathcal{M}=11$ run.

Some further caveats are commonplace ones associated with two-dimensional simulations: e.g., in reality, the binary orbital plane may be tilted with respect to the accretion disk; the vertically integrated fluid equations may not be sufficiently accurate in highly dynamical settings such as circumbinary accretion; the constant-$\alpha$ prescription for unresolved turbulence and magnetic fields may be inadequate, etc. Furthermore, we have not varied the binary mass ratio, we simulated a Newtonian system, and neglected the self-gravity of the gas. Restricting to equal-mass binaries limits our ability to inform interpretations of PG 1302-102, for example, which is a SMBH binary candidate crucially believed to have a small mass ratio, which increases the orbital velocity of the secondary black hole to account for large amplitude sinusoidal modulations via Doppler brightening \citep{dorazio1+2015}. However, we did pay special attention to our system parameters to ensure the disk is reasonably Toomre-stable, optically-thick, far from the ionization-unstable regime ($T\!\sim\!6500\,$K), not deep within the gravitational-wave driven regime (so that large eccentricity is not precluded), etc.

\section{Conclusions \& Outlook} \label{sec:conclude}
We study eccentric and circular equal-mass binaries near the transition between gas- and GW-driven evolution using two-dimensional simulations. We report multi-band light curves for the thermal disk emission, and compared them to jet emission under the assumption that jet power is proportional to the accretion rate. We find that optical and infrared (low-energy) disk emission in different bands are generally in-phase to within $\sim2\%$ of an orbital period, and that the low-energy emission is in-phase with the long term (i.e.~orbital) variability of the accretion rate for eccentric binaries. Tantalizingly, low-energy disk emission is almost completely out-of-phase with accretion rates for circular binaries, and is thus a possible smoking-gun signature of circularity. This seems consistent with the fact that circular binaries harbor larger minidisks than eccentric binaries, since one expects larger minidisks to provide more effective buffering of incoming perturbations. It is also clear that periodic low-energy disk emission can have pulse substructure that is quite distinct from accretion rates.

We find that the well-known ``lump'' period features more prominently in the low-energy disk emission from circular binaries (compared to eccentric binaries), and more prominently in emission coming from accretion streams and the cavity wall (compared to the minidisks). The lump period is virtually absent in accretion rates, suggesting that jet power could not be modulated at the lump frequency through the mechanism of mass accretion. An alternative mechanism for lump periodicity to imprint upon jet power is inverse Compton scattering of low-energy photons originating in the disk, in which case the time delays between CBD brightening and jet brightening would be much smaller than an orbital period.

We also compare the amplitude of periodic, hydrodynamic light curve variability to periodic Doppler brightening signals arising from bulk translational motion of the minidisks. We estimate that Doppler brightening is only on par with hydrodynamic variability for very high orbital velocities, typically $v/c \gtrsim 0.12-0.16$. Doppler brightening and gravitational redshift have previously been reported to primarily cause an overall dimming effect for equal-mass binaries near merger \citep{Tang2018}, rather than a smooth sinusoidal modulation. We therefore conclude that periodicity in low-energy light curves from disks around equal-mass binaries (with characteristic disk Mach numbers $\lesssim 20$) generically has a spiky, possibly noisy character associated with hydrodynamic effects, rather than the smooth and orderly character associated with Doppler brightening. We provided simulation data in Tables~\ref{tab:circ_MD} \&~\ref{tab:ecc_MD} which we hope will be useful for modeling Doppler brightening from accreting binaries.

We also found that the RMS variability of light curves is generally less than the $\sim 10\%$ stochastic variability found in typical AGN (see Tables~\ref{tab:circ_var} \&~\ref{tab:ecc_var}). Phase-folding a longer temporal baseline of data is a strategy which might reduce the stochastic component of light curves and reveal the periodic signal.

Our results here should aid the identification of genuine binaries among candidates identified as periodic quasars.  The time-domain dataset from the forthcoming Vera Rubin Observatory's Legacy Survey of Space and Time (LSST) will replace current samples (of a few $\times$ 100,000 quasars with sparsely sampled light-curves in a single band) with tens of millions of quasars sampled at the much higher cadence of once per few days in multiple bands~\citep{LSSTScienceCollaboration2009,Ivezic2019}. This will allow searches for periodic quasars at much higher fidelity and extending to much shorter periods~\citep{XinHaiman2021}, thereby providing a way to diminish stochastic variability.  Our results will be especially applicable to this large anticipated AGN-variability dataset.

\acknowledgements
We thank Geoffrey Ryan and Yan-Fei Jiang for helpful discussions. All simulations were performed on Clemson University's Palmetto cluster, and we gratefully acknowledge the Palmetto HPC support team. We acknowledge support from NSF grants AST-2006176 (to ZH) and 1715661 (to ZH and AM), and use of the software Matplotlib \citep{Hunter:2007} and SciPy \citep{scipy}.

\renewcommand\bibname{References}

\bibliographystyle{aasjournal}
\bibliography{cbd}

\begin{thebibliography}{}
\expandafter\ifx\csname natexlab\endcsname\relax\def\natexlab#1{#1}\fi
\providecommand{\url}[1]{\href{#1}{#1}}

\bibitem[{{Amaro-Seoane} {et~al.}(2017){Amaro-Seoane}, {Audley}, {Babak},
  {Baker}, {Barausse}, {Bender}, {Berti}, {Binetruy}, {Born}, {Bortoluzzi},
  {Camp}, {Caprini}, {Cardoso}, {Colpi}, {Conklin}, {Cornish}, {Cutler},
  {Danzmann}, {Dolesi}, {Ferraioli}, {Ferroni}, {Fitzsimons}, {Gair}, {Gesa
  Bote}, {Giardini}, {Gibert}, {Grimani}, {Halloin}, {Heinzel}, {Hertog},
  {Hewitson}, {Holley-Bockelmann}, {Hollington}, {Hueller}, {Inchauspe},
  {Jetzer}, {Karnesis}, {Killow}, {Klein}, {Klipstein}, {Korsakova}, {Larson},
  {Livas}, {Lloro}, {Man}, {Mance}, {Martino}, {Mateos}, {McKenzie},
  {McWilliams}, {Miller}, {Mueller}, {Nardini}, {Nelemans}, {Nofrarias},
  {Petiteau}, {Pivato}, {Plagnol}, {Porter}, {Reiche}, {Robertson},
  {Robertson}, {Rossi}, {Russano}, {Schutz}, {Sesana}, {Shoemaker}, {Slutsky},
  {Sopuerta}, {Sumner}, {Tamanini}, {Thorpe}, {Troebs}, {Vallisneri},
  {Vecchio}, {Vetrugno}, {Vitale}, {Volonteri}, {Wanner}, {Ward}, {Wass},
  {Weber}, {Ziemer}, \& {Zweifel}}]{LISA2017}
{Amaro-Seoane}, P., {Audley}, H., {Babak}, S., {et~al.} 2017, arXiv e-prints,
  arXiv:1702.00786

\bibitem[{{Armitage} \& {Natarajan}(2005)}]{Armitage2005}
{Armitage}, P.~J., \& {Natarajan}, P. 2005, \apj, 634, 921

\bibitem[{{Arzoumanian} {et~al.}(2020){Arzoumanian}, {Baker}, {Blumer},
  {B{\'e}csy}, {Brazier}, {Brook}, {Burke-Spolaor}, {Chatterjee}, {Chen},
  {Cordes}, {Cornish}, {Crawford}, {Cromartie}, {Decesar}, {Demorest}, {Dolch},
  {Ellis}, {Ferrara}, {Fiore}, {Fonseca}, {Garver-Daniels}, {Gentile}, {Good},
  {Hazboun}, {Holgado}, {Islo}, {Jennings}, {Jones}, {Kaiser}, {Kaplan},
  {Kelley}, {Key}, {Laal}, {Lam}, {Lazio}, {Lorimer}, {Luo}, {Lynch},
  {Madison}, {McLaughlin}, {Mingarelli}, {Ng}, {Nice}, {Pennucci}, {Pol},
  {Ransom}, {Ray}, {Shapiro-Albert}, {Siemens}, {Simon}, {Spiewak}, {Stairs},
  {Stinebring}, {Stovall}, {Sun}, {Swiggum}, {Taylor}, {Turner}, {Vallisneri},
  {Vigeland}, {Witt}, \& {Nanograv Collaboration}}]{PTA2020}
{Arzoumanian}, Z., {Baker}, P.~T., {Blumer}, H., {et~al.} 2020, \apjl, 905, L34

\bibitem[{{Begelman} {et~al.}(1980){Begelman}, {Blandford}, \&
  {Rees}}]{begelman+1980}
{Begelman}, M.~C., {Blandford}, R.~D., \& {Rees}, M.~J. 1980, \nat, 287, 307

\bibitem[{{Bhatta}(2021)}]{bhatta2021}
{Bhatta}, G. 2021, arXiv e-prints, arXiv:2109.08110

\bibitem[{{Bode} {et~al.}(2012){Bode}, {Bogdanovi{\'c}}, {Haas}, {Healy},
  {Laguna}, \& {Shoemaker}}]{bode+2012}
{Bode}, T., {Bogdanovi{\'c}}, T., {Haas}, R., {et~al.} 2012, \apj, 744, 45

\bibitem[{{Bogdanovi{\'c}} {et~al.}(2008){Bogdanovi{\'c}}, {Smith},
  {Sigurdsson}, \& {Eracleous}}]{bogdanovic+2008}
{Bogdanovi{\'c}}, T., {Smith}, B.~D., {Sigurdsson}, S., \& {Eracleous}, M.
  2008, \apjs, 174, 455

\bibitem[{{Charisi} {et~al.}(2016){Charisi}, {Bartos}, {Haiman},
  {Price-Whelan}, {Graham}, {Bellm}, {Laher}, \& {M{\'a}rka}}]{charisi+2016}
{Charisi}, M., {Bartos}, I., {Haiman}, Z., {et~al.} 2016, \mnras, 463, 2145

\bibitem[{{Charisi} {et~al.}(2021){Charisi}, {Taylor}, {Runnoe}, {Bogdanovic},
  \& {Trump}}]{Charisi2021}
{Charisi}, M., {Taylor}, S.~R., {Runnoe}, J., {Bogdanovic}, T., \& {Trump},
  J.~R. 2021, arXiv e-prints, arXiv:2110.14661

\bibitem[{{Chatterjee} {et~al.}(2013){Chatterjee}, {Nalewajko}, \&
  {Myers}}]{chatterjee+2013}
{Chatterjee}, R., {Nalewajko}, K., \& {Myers}, A.~D. 2013, \apjl, 771, L25

\bibitem[{{Chen} {et~al.}(2020){Chen}, {Liu}, {Liao}, {Holgado}, {Guo},
  {Gruendl}, {Morganson}, {Shen}, {Zhang}, {Abbott}, {Aguena}, {Allam},
  {Avila}, {Bertin}, {Bhargava}, {Brooks}, {Burke}, {Carnero Rosell},
  {Carollo}, {Carrasco Kind}, {Carretero}, {Costanzi}, {da Costa}, {Davis}, {De
  Vicente}, {Desai}, {Diehl}, {Doel}, {Everett}, {Flaugher}, {Friedel},
  {Frieman}, {Garc{\'\i}a-Bellido}, {Gaztanaga}, {Glazebrook}, {Gruen},
  {Gutierrez}, {Hinton}, {Hollowood}, {James}, {Kim}, {Kuehn}, {Kuropatkin},
  {Lewis}, {Lidman}, {Lima}, {Maia}, {March}, {Marshall}, {Menanteau},
  {Miquel}, {Palmese}, {Paz-Chinch{\'o}n}, {Plazas}, {Sanchez}, {Schubnell},
  {Serrano}, {Sevilla-Noarbe}, {Smith}, {Suchyta}, {Swanson}, {Tarle},
  {Tucker}, {Norbert Varga}, \& {Walker}}]{chen+2020}
{Chen}, Y.-C., {Liu}, X., {Liao}, W.-T., {et~al.} 2020, \mnras, 499, 2245

\bibitem[{{Combi} {et~al.}(2021){Combi}, {Lopez Armengol}, {Campanelli},
  {Noble}, {Avara}, {Krolik}, \& {Bowen}}]{combi+2021}
{Combi}, L., {Lopez Armengol}, F.~G., {Campanelli}, M., {et~al.} 2021, arXiv
  e-prints, arXiv:2109.01307

\bibitem[{{Covino} {et~al.}(2020){Covino}, {Landoni}, {Sandrinelli}, \&
  {Treves}}]{covino+2020}
{Covino}, S., {Landoni}, M., {Sandrinelli}, A., \& {Treves}, A. 2020, \apj,
  895, 122

\bibitem[{{d'Ascoli} {et~al.}(2018){d'Ascoli}, {Noble}, {Bowen}, {Campanelli},
  {Krolik}, \& {Mewes}}]{dAscoli+2018}
{d'Ascoli}, S., {Noble}, S.~C., {Bowen}, D.~B., {et~al.} 2018, \apj, 865, 140

\bibitem[{{De Rosa} {et~al.}(2019){De Rosa}, {Vignali}, {Bogdanovi{\'c}},
  {Capelo}, {Charisi}, {Dotti}, {Husemann}, {Lusso}, {Mayer}, {Paragi},
  {Runnoe}, {Sesana}, {Steinborn}, {Bianchi}, {Colpi}, {del Valle}, {Frey},
  {Gab{\'a}nyi}, {Giustini}, {Guainazzi}, {Haiman}, {Herrera Ruiz},
  {Herrero-Illana}, {Iwasawa}, {Komossa}, {Lena}, {Loiseau}, {Perez-Torres},
  {Piconcelli}, \& {Volonteri}}]{derosa+2019}
{De Rosa}, A., {Vignali}, C., {Bogdanovi{\'c}}, T., {et~al.} 2019, \nar, 86,
  101525

\bibitem[{{Dempsey} {et~al.}(2020){Dempsey}, {Mu{\~n}oz}, \&
  {Lithwick}}]{dempsey+2020}
{Dempsey}, A.~M., {Mu{\~n}oz}, D., \& {Lithwick}, Y. 2020, \apjl, 892, L29

\bibitem[{{Dittmann} \& {Ryan}(2021)}]{Dittmann+2021}
{Dittmann}, A., \& {Ryan}, G. 2021, arXiv e-prints, arXiv:2102.05684

\bibitem[{{D'Orazio} \& {Duffell}(2021)}]{DOrazio:2021:eccentric}
{D'Orazio}, D.~J., \& {Duffell}, P.~C. 2021, arXiv e-prints, arXiv:2103.09251

\bibitem[{{D'Orazio} {et~al.}(2016){D'Orazio}, {Haiman}, {Duffell},
  {MacFadyen}, \& {Farris}}]{Dorazio+2016}
{D'Orazio}, D.~J., {Haiman}, Z., {Duffell}, P., {MacFadyen}, A., \& {Farris},
  B. 2016, \mnras, 459, 2379

\bibitem[{{D'Orazio} {et~al.}(2015){D'Orazio}, {Haiman}, \&
  {Schiminovich}}]{dorazio1+2015}
{D'Orazio}, D.~J., {Haiman}, Z., \& {Schiminovich}, D. 2015, \nat, 525, 351

\bibitem[{{Duffell} {et~al.}(2020){Duffell}, {D'Orazio}, {Derdzinski},
  {Haiman}, {MacFadyen}, {Rosen}, \& {Zrake}}]{duffel:2020:massratio}
{Duffell}, P.~C., {D'Orazio}, D., {Derdzinski}, A., {et~al.} 2020, \apj, 901,
  25

\bibitem[{{Edelson} \& {Krolik}(1988)}]{Edelson+1988}
{Edelson}, R.~A., \& {Krolik}, J.~H. 1988, \apj, 333, 646

\bibitem[{{Eggleton}(1983)}]{eggleton1983}
{Eggleton}, P.~P. 1983, \apj, 268, 368

\bibitem[{{Farris} {et~al.}(2014){Farris}, {Duffell}, {MacFadyen}, \&
  {Haiman}}]{Farris+2014}
{Farris}, B.~D., {Duffell}, P., {MacFadyen}, A.~I., \& {Haiman}, Z. 2014, \apj,
  783, 134

\bibitem[{{Farris} {et~al.}(2015{\natexlab{a}}){Farris}, {Duffell},
  {MacFadyen}, \& {Haiman}}]{farris+2015b}
---. 2015{\natexlab{a}}, \mnras, 446, L36

\bibitem[{{Farris} {et~al.}(2015{\natexlab{b}}){Farris}, {Duffell},
  {MacFadyen}, \& {Haiman}}]{farris+2015}
---. 2015{\natexlab{b}}, \mnras, 447, L80

\bibitem[{{Ferrarese} \& {Ford}(2005)}]{ferrarese+2005}
{Ferrarese}, L., \& {Ford}, H. 2005, \ssr, 116, 523

\bibitem[{{Frank} {et~al.}(2002){Frank}, {King}, \& {Raine}}]{FKR}
{Frank}, J., {King}, A., \& {Raine}, D.~J. 2002, {Accretion Power in
  Astrophysics: Third Edition}

\bibitem[{{Giacomazzo} {et~al.}(2012){Giacomazzo}, {Baker}, {Miller},
  {Reynolds}, \& {van Meter}}]{giacomazzo+2012}
{Giacomazzo}, B., {Baker}, J.~G., {Miller}, M.~C., {Reynolds}, C.~S., \& {van
  Meter}, J.~R. 2012, \apjl, 752, L15

\bibitem[{{Gold} {et~al.}(2014){Gold}, {Paschalidis}, {Ruiz}, {Shapiro},
  {Etienne}, \& {Pfeiffer}}]{gold2014}
{Gold}, R., {Paschalidis}, V., {Ruiz}, M., {et~al.} 2014, \prd, 90, 104030

\bibitem[{{Goodman}(2003)}]{Goodman2003}
{Goodman}, J. 2003, \mnras, 339, 937

\bibitem[{{Graham} {et~al.}(2015){Graham}, {Djorgovski}, {Stern}, {Drake},
  {Mahabal}, {Donalek}, {Glikman}, {Larson}, \& {Christensen}}]{graham+2015}
{Graham}, M.~J., {Djorgovski}, S.~G., {Stern}, D., {et~al.} 2015, \mnras, 453,
  1562

\bibitem[{{Guti{\'e}rrez} {et~al.}(2021){Guti{\'e}rrez}, {Combi}, {Noble},
  {Campanelli}, {Krolik}, {L{\'o}pez Armengol}, \&
  {Garc{\'\i}a}}]{Gutierrez+2021}
{Guti{\'e}rrez}, E.~M., {Combi}, L., {Noble}, S.~C., {et~al.} 2021, arXiv
  e-prints, arXiv:2112.09773

\bibitem[{{Hu} {et~al.}(2020){Hu}, {D'Orazio}, {Haiman}, {Smith}, {Snios},
  {Charisi}, \& {Di Stefano}}]{hu+2020}
{Hu}, B.~X., {D'Orazio}, D.~J., {Haiman}, Z., {et~al.} 2020, \mnras, 495, 4061

\bibitem[{Hunter(2007)}]{Hunter:2007}
Hunter, J.~D. 2007, Computing in Science \& Engineering, 9, 90

\bibitem[{{Ivezi{\'c}} {et~al.}(2019){Ivezi{\'c}}, {Kahn}, {Tyson}, {Abel},
  {Acosta}, {Allsman}, {Alonso}, {AlSayyad}, {Anderson}, {Andrew}, {Angel},
  {Angeli}, {Ansari}, {Antilogus}, {Araujo}, {Armstrong}, {Arndt}, {Astier},
  {Aubourg}, {Auza}, {Axelrod}, {Bard}, {Barr}, {Barrau}, {Bartlett}, {Bauer},
  {Bauman}, {Baumont}, {Bechtol}, {Bechtol}, {Becker}, {Becla}, {Beldica},
  {Bellavia}, {Bianco}, {Biswas}, {Blanc}, {Blazek}, {Blandford}, {Bloom},
  {Bogart}, {Bond}, {Booth}, {Borgland}, {Borne}, {Bosch}, {Boutigny},
  {Brackett}, {Bradshaw}, {Brandt}, {Brown}, {Bullock}, {Burchat}, {Burke},
  {Cagnoli}, {Calabrese}, {Callahan}, {Callen}, {Carlin}, {Carlson},
  {Chandrasekharan}, {Charles-Emerson}, {Chesley}, {Cheu}, {Chiang}, {Chiang},
  {Chirino}, {Chow}, {Ciardi}, {Claver}, {Cohen-Tanugi}, {Cockrum}, {Coles},
  {Connolly}, {Cook}, {Cooray}, {Covey}, {Cribbs}, {Cui}, {Cutri}, {Daly},
  {Daniel}, {Daruich}, {Daubard}, {Daues}, {Dawson}, {Delgado}, {Dellapenna},
  {de Peyster}, {de Val-Borro}, {Digel}, {Doherty}, {Dubois},
  {Dubois-Felsmann}, {Durech}, {Economou}, {Eifler}, {Eracleous}, {Emmons},
  {Fausti Neto}, {Ferguson}, {Figueroa}, {Fisher-Levine}, {Focke}, {Foss},
  {Frank}, {Freemon}, {Gangler}, {Gawiser}, {Geary}, {Gee}, {Geha}, {Gessner},
  {Gibson}, {Gilmore}, {Glanzman}, {Glick}, {Goldina}, {Goldstein}, {Goodenow},
  {Graham}, {Gressler}, {Gris}, {Guy}, {Guyonnet}, {Haller}, {Harris},
  {Hascall}, {Haupt}, {Hernandez}, {Herrmann}, {Hileman}, {Hoblitt}, {Hodgson},
  {Hogan}, {Howard}, {Huang}, {Huffer}, {Ingraham}, {Innes}, {Jacoby}, {Jain},
  {Jammes}, {Jee}, {Jenness}, {Jernigan}, {Jevremovi{\'c}}, {Johns}, {Johnson},
  {Johnson}, {Jones}, {Juramy-Gilles}, {Juri{\'c}}, {Kalirai}, {Kallivayalil},
  {Kalmbach}, {Kantor}, {Karst}, {Kasliwal}, {Kelly}, {Kessler}, {Kinnison},
  {Kirkby}, {Knox}, {Kotov}, {Krabbendam}, {Krughoff}, {Kub{\'a}nek},
  {Kuczewski}, {Kulkarni}, {Ku}, {Kurita}, {Lage}, {Lambert}, {Lange},
  {Langton}, {Le Guillou}, {Levine}, {Liang}, {Lim}, {Lintott}, {Long},
  {Lopez}, {Lotz}, {Lupton}, {Lust}, {MacArthur}, {Mahabal}, {Mandelbaum},
  {Markiewicz}, {Marsh}, {Marshall}, {Marshall}, {May}, {McKercher}, {McQueen},
  {Meyers}, {Migliore}, {Miller}, {Mills}, {Miraval}, {Moeyens}, {Moolekamp},
  {Monet}, {Moniez}, {Monkewitz}, {Montgomery}, {Morrison}, {Mueller},
  {Muller}, {Mu{\~n}oz Arancibia}, {Neill}, {Newbry}, {Nief}, {Nomerotski},
  {Nordby}, {O'Connor}, {Oliver}, {Olivier}, {Olsen}, {O'Mullane}, {Ortiz},
  {Osier}, {Owen}, {Pain}, {Palecek}, {Parejko}, {Parsons}, {Pease},
  {Peterson}, {Peterson}, {Petravick}, {Libby Petrick}, {Petry},
  {Pierfederici}, {Pietrowicz}, {Pike}, {Pinto}, {Plante}, {Plate}, {Plutchak},
  {Price}, {Prouza}, {Radeka}, {Rajagopal}, {Rasmussen}, {Regnault}, {Reil},
  {Reiss}, {Reuter}, {Ridgway}, {Riot}, {Ritz}, {Robinson}, {Roby}, {Roodman},
  {Rosing}, {Roucelle}, {Rumore}, {Russo}, {Saha}, {Sassolas}, {Schalk},
  {Schellart}, {Schindler}, {Schmidt}, {Schneider}, {Schneider}, {Schoening},
  {Schumacher}, {Schwamb}, {Sebag}, {Selvy}, {Sembroski}, {Seppala}, {Serio},
  {Serrano}, {Shaw}, {Shipsey}, {Sick}, {Silvestri}, {Slater}, {Smith},
  {Smith}, {Sobhani}, {Soldahl}, {Storrie-Lombardi}, {Stover}, {Strauss},
  {Street}, {Stubbs}, {Sullivan}, {Sweeney}, {Swinbank}, {Szalay}, {Takacs},
  {Tether}, {Thaler}, {Thayer}, {Thomas}, {Thornton}, {Thukral}, {Tice},
  {Trilling}, {Turri}, {Van Berg}, {Vanden Berk}, {Vetter}, {Virieux},
  {Vucina}, {Wahl}, {Walkowicz}, {Walsh}, {Walter}, {Wang}, {Wang}, {Warner},
  {Wiecha}, {Willman}, {Winters}, {Wittman}, {Wolff}, {Wood-Vasey}, {Wu},
  {Xin}, {Yoachim}, \& {Zhan}}]{Ivezic2019}
{Ivezi{\'c}}, {\v{Z}}., {Kahn}, S.~M., {Tyson}, J.~A., {et~al.} 2019, \apj,
  873, 111

\bibitem[{{Jiang} \& {Blaes}(2020)}]{jiang+2020}
{Jiang}, Y.-F., \& {Blaes}, O. 2020, \apj, 900, 25

\bibitem[{{Jiang} {et~al.}(2019){Jiang}, {Stone}, \& {Davis}}]{Jiang+2019}
{Jiang}, Y.-F., {Stone}, J.~M., \& {Davis}, S.~W. 2019, \apj, 880, 67

\bibitem[{Jones {et~al.}(2001--)Jones, Oliphant, Peterson, {et~al.}}]{scipy}
Jones, E., Oliphant, T., Peterson, P., {et~al.} 2001--, {SciPy}: Open source
  scientific tools for {Python}, , , [Online; accessed 2019-06-01].
\newblock \url{http://www.scipy.org/}

\bibitem[{{Khatoon} {et~al.}(2021){Khatoon}, {Prince}, {Shah}, {Sahayanathan},
  \& {Gogoi}}]{khatoon+2021}
{Khatoon}, R., {Prince}, R., {Shah}, Z., {Sahayanathan}, S., \& {Gogoi}, R.
  2021, arXiv e-prints, arXiv:2104.12130

\bibitem[{{Kormendy} \& {Ho}(2013)}]{kormendy+2013}
{Kormendy}, J., \& {Ho}, L.~C. 2013, \araa, 51, 511

\bibitem[{{Kormendy} \& {Richstone}(1995)}]{kormendy+1995}
{Kormendy}, J., \& {Richstone}, D. 1995, \araa, 33, 581

\bibitem[{{Lightman} \& {Eardley}(1974)}]{lightman+1974}
{Lightman}, A.~P., \& {Eardley}, D.~M. 1974, \apjl, 187, L1

\bibitem[{{Liu} {et~al.}(2019){Liu}, {Gezari}, {Ayers}, {Burgett}, {Chambers},
  {Hodapp}, {Huber}, {Kudritzki}, {Metcalfe}, {Tonry}, {Wainscoat}, \&
  {Waters}}]{liu+2019}
{Liu}, T., {Gezari}, S., {Ayers}, M., {et~al.} 2019, \apj, 884, 36

\bibitem[{{Liu} {et~al.}(2020){Liu}, {Koss}, {Blecha}, {Ricci}, {Trakhtenbrot},
  {Mushotzky}, {Harrison}, {Ichikawa}, {Kakkad}, {Oh}, {Powell}, {Privon},
  {Schawinski}, {Shimizu}, {Smith}, {Stern}, {Treister}, \& {Urry}}]{liu+2020}
{Liu}, T., {Koss}, M., {Blecha}, L., {et~al.} 2020, \apj, 896, 122

\bibitem[{{Lotz} {et~al.}(2011){Lotz}, {Jonsson}, {Cox}, {Croton}, {Primack},
  {Somerville}, \& {Stewart}}]{lotz+2011}
{Lotz}, J.~M., {Jonsson}, P., {Cox}, T.~J., {et~al.} 2011, \apj, 742, 103

\bibitem[{{LSST Science Collaboration} {et~al.}(2009){LSST Science
  Collaboration}, Abell, Allison, Anderson, Andrew, Angel, Armus, Arnett,
  Asztalos, Axelrod, Bailey, Ballantyne, Bankert, Barkhouse, Barr, Barrientos,
  Barth, Bartlett, Becker, Becla, Beers, Bernstein, Biswas, Blanton, Bloom,
  Bochanski, Boeshaar, Borne, Bradac, Brandt, Bridge, Brown, Brunner, Bullock,
  Burgasser, Burge, Burke, Cargile, Chandrasekharan, Chartas, Chesley, Chu,
  Cinabro, Claire, Claver, Clowe, Connolly, Cook, Cooke, Cooray, Covey,
  Culliton, de~Jong, de~Vries, Debattista, Delgado, Dell'Antonio, Dhital, {Di
  Stefano}, Dickinson, Dilday, Djorgovski, Dobler, Donalek, Dubois-Felsmann,
  Durech, Eliasdottir, Eracleous, Eyer, Falco, Fan, Fassnacht, Ferguson,
  Fernandez, Fields, Finkbeiner, Figueroa, Fox, Francke, Frank, Frieman,
  Fromenteau, Furqan, Galaz, Gal-Yam, Garnavich, Gawiser, Geary, Gee, Gibson,
  Gilmore, Grace, Green, Gressler, Grillmair, Habib, Haggerty, Hamuy, Harris,
  Hawley, Heavens, Hebb, Henry, Hileman, Hilton, Hoadley, Holberg, Holman,
  Howell, Infante, Ivezic, Jacoby, Jain, R, Jedicke, Jee, Jernigan, Jha,
  Johnston, Jones, Juric, Kaasalainen, Styliani, Kafka, Kahn, Kaib, Kalirai,
  Kantor, Kasliwal, Keeton, Kessler, Knezevic, Kowalski, Krabbendam, Krughoff,
  Kulkarni, Kuhlman, Lacy, Lepine, Liang, Lien, Lira, Long, Lorenz, Lotz,
  Lupton, Lutz, Macri, Mahabal, Mandelbaum, Marshall, May, McGehee, Meadows,
  Meert, Milani, Miller, Miller, Mills, Minniti, Monet, Mukadam, Nakar, Neill,
  Newman, Nikolaev, Nordby, O'Connor, Oguri, Oliver, Olivier, Olsen, Olsen,
  Olszewski, Oluseyi, Padilla, Parker, Pepper, Peterson, Petry, Pinto, Pizagno,
  Popescu, Prsa, Radcka, Raddick, Rasmussen, Rau, Rho, Rhoads, Richards,
  Ridgway, Robertson, Roskar, Saha, Sarajedini, Scannapieco, Schalk, Schindler,
  Schmidt, Schmidt, Schneider, Schumacher, Scranton, Sebag, Seppala, Shemmer,
  Simon, Sivertz, Smith, Smith, Smith, Spitz, Stanford, Stassun, Strader,
  Strauss, Stubbs, Sweeney, Szalay, Szkody, Takada, Thorman, Trilling, Trimble,
  Tyson, {Van Berg}, Berk, VanderPlas, Verde, Vrsnak, Walkowicz, Wandelt, Wang,
  Wang, Warner, Wechsler, West, Wiecha, Williams, Willman, Wittman, Wolff,
  Wood-Vasey, Wozniak, Young, Zentner, \& Zhan}]{LSSTScienceCollaboration2009}
{LSST Science Collaboration}, Abell, P.~A., Allison, J., {et~al.} 2009, arXiv
  e-prints, arXiv:0912.0201.
\newblock \url{http://arxiv.org/abs/0912.0201}

\bibitem[{{Negi} {et~al.}(2021){Negi}, {Joshi}, {Chand}, {Chand}, {Wiita},
  {Ho}, \& {Singh}}]{Negi+2021}
{Negi}, V., {Joshi}, R., {Chand}, K., {et~al.} 2021, arXiv e-prints,
  arXiv:2112.00790

\bibitem[{{Noble} {et~al.}(2012){Noble}, {Mundim}, {Nakano}, {Krolik},
  {Campanelli}, {Zlochower}, \& {Yunes}}]{noble+2012}
{Noble}, S.~C., {Mundim}, B.~C., {Nakano}, H., {et~al.} 2012, \apj, 755, 51

\bibitem[{{Paczynski}(1991)}]{paczynski1991}
{Paczynski}, B. 1991, \apj, 370, 597

\bibitem[{{Padovani} {et~al.}(2017){Padovani}, {Alexander}, {Assef}, {De
  Marco}, {Giommi}, {Hickox}, {Richards}, {Smol{\v{c}}i{\'c}},
  {Hatziminaoglou}, {Mainieri}, \& {Salvato}}]{padovani+2017}
{Padovani}, P., {Alexander}, D.~M., {Assef}, R.~J., {et~al.} 2017, \aapr, 25, 2

\bibitem[{{Paschalidis} {et~al.}(2021){Paschalidis}, {Bright}, {Ruiz}, \&
  {Gold}}]{paschalidis+2021}
{Paschalidis}, V., {Bright}, J., {Ruiz}, M., \& {Gold}, R. 2021, \apjl, 910,
  L26

\bibitem[{{Pe{\~n}il} {et~al.}(2020){Pe{\~n}il}, {Dom{\'\i}nguez}, {Buson},
  {Ajello}, {Otero-Santos}, {Barrio}, {Nemmen}, {Cutini}, {Rani},
  {Franckowiak}, \& {Cavazzuti}}]{penil+2020}
{Pe{\~n}il}, P., {Dom{\'\i}nguez}, A., {Buson}, S., {et~al.} 2020, \apj, 896,
  134

\bibitem[{{Peters}(1964)}]{peters1964}
{Peters}, P.~C. 1964, Physical Review, 136, 1224

\bibitem[{{Peterson} {et~al.}(1976){Peterson}, {Jauncey}, {Wright}, \&
  {Condon}}]{0208redshift}
{Peterson}, B.~A., {Jauncey}, D.~J., {Wright}, A.~E., \& {Condon}, J.~J. 1976,
  \apjl, 207, L5

\bibitem[{{Popham} \& {Narayan}(1991)}]{popham+1991}
{Popham}, R., \& {Narayan}, R. 1991, \apj, 370, 604

\bibitem[{{Roedig} {et~al.}(2011){Roedig}, {Dotti}, {Sesana}, {Cuadra}, \&
  {Colpi}}]{Roedig+2011}
{Roedig}, C., {Dotti}, M., {Sesana}, A., {Cuadra}, J., \& {Colpi}, M. 2011,
  \mnras, 415, 3033

\bibitem[{{Roedig} \& {Sesana}(2014)}]{Roedig+2014}
{Roedig}, C., \& {Sesana}, A. 2014, \mnras, 439, 3476

\bibitem[{{Roedig} {et~al.}(2012){Roedig}, {Sesana}, {Dotti}, {Cuadra},
  {Amaro-Seoane}, \& {Haardt}}]{Roedig+2012}
{Roedig}, C., {Sesana}, A., {Dotti}, M., {et~al.} 2012, \aap, 545, A127

\bibitem[{{Ryan} \& {MacFadyen}(2017)}]{Ryan+2017}
{Ryan}, G., \& {MacFadyen}, A. 2017, \apj, 835, 199

\bibitem[{{Sandrinelli} {et~al.}(2016){Sandrinelli}, {Covino}, {Dotti}, \&
  {Treves}}]{sandrinelli+2016}
{Sandrinelli}, A., {Covino}, S., {Dotti}, M., \& {Treves}, A. 2016, \aj, 151,
  54

\bibitem[{{Sandrinelli} {et~al.}(2014){Sandrinelli}, {Covino}, \&
  {Treves}}]{sandrinelli+2014}
{Sandrinelli}, A., {Covino}, S., \& {Treves}, A. 2014, \apjl, 793, L1

\bibitem[{{Shakura} \& {Sunyaev}(1973)}]{SS1973}
{Shakura}, N.~I., \& {Sunyaev}, R.~A. 1973, \aap, 500, 33

\bibitem[{{Shi} \& {Krolik}(2015)}]{shi+2015}
{Shi}, J.-M., \& {Krolik}, J.~H. 2015, \apj, 807, 131

\bibitem[{{Stacy} {et~al.}(2003){Stacy}, {Vestrand}, \&
  {Sreekumar}}]{stacy0208mass}
{Stacy}, J.~G., {Vestrand}, W.~T., \& {Sreekumar}, P. 2003, \apj, 598, 216

\bibitem[{{Tang} {et~al.}(2018){Tang}, {Haiman}, \& {MacFadyen}}]{Tang2018}
{Tang}, Y., {Haiman}, Z., \& {MacFadyen}, A. 2018, \mnras, 476, 2249

\bibitem[{{Tiede} {et~al.}(2020){Tiede}, {Zrake}, {MacFadyen}, \&
  {Haiman}}]{Tiede2020}
{Tiede}, C., {Zrake}, J., {MacFadyen}, A., \& {Haiman}, Z. 2020, \apj, 900, 43

\bibitem[{{Vanden Berk} {et~al.}(2001){Vanden Berk}, {Richards}, {Bauer},
  {Strauss}, {Schneider}, {Heckman}, {York}, {Hall}, {Fan}, {Knapp},
  {Anderson}, {Annis}, {Bahcall}, {Bernardi}, {Briggs}, {Brinkmann}, {Brunner},
  {Burles}, {Carey}, {Castander}, {Connolly}, {Crocker}, {Csabai}, {Doi},
  {Finkbeiner}, {Friedman}, {Frieman}, {Fukugita}, {Gunn}, {Hennessy},
  {Ivezi{\'c}}, {Kent}, {Kunszt}, {Lamb}, {Leger}, {Long}, {Loveday}, {Lupton},
  {Meiksin}, {Merelli}, {Munn}, {Newberg}, {Newcomb}, {Nichol}, {Owen}, {Pier},
  {Pope}, {Rockosi}, {Schlegel}, {Siegmund}, {Smee}, {Snir}, {Stoughton},
  {Stubbs}, {SubbaRao}, {Szalay}, {Szokoly}, {Tremonti}, {Uomoto}, {Waddell},
  {Yanny}, \& {Zheng}}]{berk+2001}
{Vanden Berk}, D.~E., {Richards}, G.~T., {Bauer}, A., {et~al.} 2001, \aj, 122,
  549

\bibitem[{{Vaughan} {et~al.}(2016){Vaughan}, {Uttley}, {Markowitz},
  {Huppenkothen}, {Middleton}, {Alston}, {Scargle}, \& {Farr}}]{Vaughan+2016}
{Vaughan}, S., {Uttley}, P., {Markowitz}, A.~G., {et~al.} 2016, \mnras, 461,
  3145

\bibitem[{{White} \& {Rees}(1978)}]{white+1978}
{White}, S.~D.~M., \& {Rees}, M.~J. 1978, \mnras, 183, 341

\bibitem[{{Widger} \& {Woodall}(1976)}]{integrateplanck}
{Widger}, W.~K., J., \& {Woodall}, M.~P. 1976, Bulletin of the American
  Meteorological Society, 57, 1217

\bibitem[{{Xin} \& {Haiman}(2021)}]{XinHaiman2021}
{Xin}, C., \& {Haiman}, Z. 2021, \mnras, 506, 2408

\bibitem[{{Yang} \& {Fan}(2010)}]{yang0208mass}
{Yang}, J., \& {Fan}, J. 2010, Science China Physics, Mechanics, and Astronomy,
  53, 1921

\bibitem[{{Zhang} {et~al.}(2014){Zhang}, {Zhao}, {Wang}, \& {Dai}}]{zhang+2014}
{Zhang}, B.-K., {Zhao}, X.-Y., {Wang}, C.-X., \& {Dai}, B.-Z. 2014, Research in
  Astronomy and Astrophysics, 14, 933

\bibitem[{{Zrake} {et~al.}(2021){Zrake}, {Tiede}, {MacFadyen}, \&
  {Haiman}}]{Zrake:2021:eccentric}
{Zrake}, J., {Tiede}, C., {MacFadyen}, A., \& {Haiman}, Z. 2021, \apjl, 909,
  L13

\end{thebibliography}

\appendix
%
%
\section{Numerical prescriptions} \label{app:fixes}
We use density and pressure floors in the conservative-to-primitive variable transformation. The floor values are respectively set to $\Sigma(t\!=\!0,r\!=\!a)\times 10^{-10}$ \& $\mathcal{P}(t\!=\!0,r\!=\!a)\times 10^{-10}$. If the pressure is found to be below the floor value, it is set to the floor value. If the density is found to be below the floor value, both the density and pressure are set to their floor values and the velocity is set to zero.

A component-wise velocity ceiling is applied in the conserved-to-primitive variable transformation. If a velocity component is found to be greater than $10a\Omega_{\rm bin}$, then it is set to $10a\Omega_{\rm bin}$. This compares with a typical maximum speed on the grid of $\simeq 3a\Omega_{\rm bin}$, corresponding to the orbital speed at the sink radius. We find that the velocity ceiling is rarely invoked, but it prevents the time step from becoming prohibitively small in rare scenarios whereby a fluid element is temporarily accelerated to high speeds in the neighborhood of a sink. For reference, the speed of light is approximately $50a\Omega_{\rm bin}$ for our binary parameters.

We employ a Mach ceiling to the cooling prescription, which acts to regulate the cooling strength when the updated Mach number would exceed the Mach ceiling $\mathcal{M}_c=10^5$. This is achieved by limiting the rate of cooling such that the updated specific internal energy (\cite{Ryan+2017}) has a minimum value of
\begin{eqnarray}
    \frac{2}{\Gamma (\Gamma-1)} \frac{v^2}{\mathcal{M}_c^2}.
\end{eqnarray}
We find the Mach ceiling is applied very infrequently.
%
%
%
\section{Sensitivity tests} \label{app:sens}
We checked the sensitivity of our results to variations in many parameters and prescriptions. The tests and their labels are enumerated below, and the labels are used when quoting results in Tables~\ref{tab:circ_var}-\ref{tab:ecc_MD}. Tabulated results are rounded to 2 significant figures. All tests except the variation of viscosity are alternate representations of our target system described in \sref{sec:binaries} \& \sref{sec:disk}. All tests are variations of our ``fiducial'' run parameters: $\mathcal{M}(1.5a)\simeq 11$, $e\in \lbrace 0, 0.45 \rbrace$, $\Delta x = 0.01a$, $D=15a$, floors 10 orders of magnitude below the initial conditions at $r=a$, $r_{\rm sink}/\Delta x = 4$, $s=10$, analysis time of 700$-$800 orbits, $r_s = r_{\rm sink}$, $\alpha=0.1$, and disk initial conditions given in \sref{sec:init}.
\begin{enumerate}
    \item \{label: $\mathcal{M}$\} Mach number $\mathcal{M}(3a) \simeq 21$, $\mathcal{M}(a) \simeq 7$ (fiducial is $\mathcal{M}(1.5a) \simeq 11$)
    \item \{label: $D$\} Domain radius: $D=20 a$ at fixed resolution $\Delta x = 0.01 a$ (fiducial is $D=15 a$)
    \item \{label: $\Delta x$\} Resolution: $\Delta x = 0.005 a$ (fiducial is $\Delta x = 0.01 a$)
    \item \{label: VC\} Velocity ceiling of $20 a \Omega_{\rm bin}$ (fiducial is $10 a \Omega_{\rm bin}$)
    \item \{label: F\} Floors on pressure \& density: 15 orders of magnitude smaller than the initial conditions at $r=a$ (fiducial is 10 orders of magnitude)
    \item \{label: $r_{\rm sink}$\} Sink radius: $r_{\rm sink} \in \lbrace 0.02 a, 0.01 a \rbrace$, keeping $r_{\rm sink}/\Delta x = 4$ constant (fiducial is $r_{\rm sink}=0.04 a$)
    \item \{label: $s$\} Sink rate: $s \in \lbrace 2 , 50 \rbrace$ (fiducial is $s=10$)
    \item \{label: AT\} Analysis time: 1300 orbits  (fiducial is 700 orbits)
    \item \{label: ET\} Evolution time from grid refinement at 600 orbits until the analysis time: 200 orbits (fiducial is 100 orbits), corresponding to analysis times 800$-$900 orbits (fiducial is 700$-$800 orbits)
    \item \{label: $\alpha$\} Viscosity: $\alpha=0.02$ (fiducial is $\alpha=0.1$)
    \item \{label: $e=0.7$\} Eccentricity of $e=0.7$ (fiducial eccentric run is $e=0.45$)
    \item Gravitational softening length $r_s$: this is changed simultaneously with the sink radius, since we fix $r_s = r_{\rm sink}$
\end{enumerate}
The disk initial conditions we use for the viscosity test are fixed to the $\alpha=0.1$ case, for simplicity; for $\alpha = 0.02$, the self-consistent initial densities and pressures at $r=a$ would vary by a factor of a few above the value in the $\alpha=0.1$ case.


\begin{table*}[t]
\begin{tabular}{|l|l|l|l|l|l|l|l|l|l|l||l|}
\hline
                                & $\mathcal{M}$ & $D$ & $\Delta x$ & VC  & F   & $r_{\rm sink}$    & $s$     & AT  & ET  & $\alpha$ & total range \\ \hline
optical avg.~[$10^{42}$ erg/s]  & 8.6-14       & 9.4 & 9.9        & 9.5 & 9.5 & 8.2-8.5                                            & 8.7-11  & 8.9 & 9.0 & 6.0      & {\bf 6.0-14 }     \\ \hline
infrared avg.~[$10^{42}$ erg/s] & 9.5-14       & 10  & 9.6        & 10  & 10  & 9.6-9.7                                            & 10-11   & 9.5 & 10  & 5.8      & {\bf 5.8-14 }     \\ \hline
optical variability [RMS \%]        & 3.3-7.7       & 5.6 & 3.3        & 5.5  & 5.5  & 5.2                                         & 5.5-5.6 & 5.9 & 5.7 & 8.5       & {\bf 3.3-8.5 }     \\ \hline
infrared variability [RMS \%]       & 1.1-2.4       & 1.4 & 0.90       & 1.4 & 1.4 & 0.95-1.1                                          & 1.4-1.5 & 1.7 & 1.4 & 2.6      & {\bf 0.90-2.6 }     \\ \hline
\end{tabular}
\caption{Circular binary: root-mean-square variability (RMS) for the fast periodic modulation $1/f_{\rm fast}$. The median peak-to-trough difference is roughly $3\times$ the RMS variability. We also provide the average luminosity.}\label{tab:circ_var}
\end{table*}

\begin{table*}[t]
\begin{tabular}{|l|l|l|l|l|l|l|l|l|l|l|l||l|}
\hline
                                & $\mathcal{M}$ & $D$ & $\Delta x$ & VC  & F   & $r_{\rm sink}$ & $s$     & AT  & ET  & $\alpha$ & $e=0.7$ & total range \\ \hline
optical avg.~[$10^{42}$ erg/s]  & 4.4-8.7       & 6.2 & 5.4        & 6.3 & 4.4 & 4.1-5.2                                        & 5.4-6.8 & 5.3 & 5.7 & 2.7      & 5.7     & {\bf 2.7-8.7 }     \\ \hline
infrared avg.~[$10^{42}$ erg/s] & 8.5-12        & 9.4 & 8.5        & 9.2 & 8.5 & 8.5-9.0                                        & 8.9-11  & 8.8 & 9.3 & 4.5      & 8.7     & {\bf 4.5-12  }     \\ \hline
optical variability [RMS \%]        & 8.4-15        & 7.7 & 9.1        & 13  & 15  & 9.5-10               &                                  13-15   & 11  & 9.7 & 15       & 13      & {\bf 7.7-15 }     \\ \hline
infrared variability [RMS \%]       & 2.0-2.6       & 2.4 & 2.3        & 3.5 & 2.8  & 1.9-2.4               &                                 3.0-3.2 & 2.6 & 2.3 & 2.2      & 3.1     & {\bf 1.9-3.5 }     \\ \hline
\end{tabular}
\caption{Eccentric binary: root-mean-square variability (RMS) for the orbital modulation. The median peak-to-trough difference is roughly $3\times$ the RMS variability. All columns are for $e=0.45$ except the column labeled $e=0.7$. We also provide the average luminosity.}\label{tab:ecc_var}
\end{table*}


\begin{table*}[t]
\begin{tabular}{|l|l|l|l|l|l|l|l|l|l|l||l|}
\hline
                                & $\mathcal{M}$ & $D$  & $\Delta x$ & VC  & F    & $r_{\rm sink}$  & $s$ & AT & ET & $\alpha$ & total range \\ \hline
$f_{\rm lump}$ [orbits$^{-1}$]  & 0.11-0.22      & 0.11 & 0.11      & 0.11 & 0.11 & 0.10-0.11                                     & 0.10-0.11 & 0.093 & 0.10 & 0.10 & {\bf 0.093-0.22 }     \\ \hline
$f_{\rm fast}$ [orbits$^{-1}$]  & 1.3-1.4        & 1.4  & 1.3       & 1.4  & 1.4  & 1.0-1.2                                       & 1.4       & 1.4   & 1.4  & 1.5 & {\bf 1.0-1.5 }     \\ \hline
\end{tabular}
\caption{Circular binary: $f_{\rm lump}$ and $f_{\rm fast}$.} \label{tab:circ_freq}
\end{table*}


\begin{table*}[t]
\begin{tabular}{|l|l|l|l|l|l|l|l|l|l|l||l|}
\hline
            & $\mathcal{M}$ & $D$  & $\Delta x$ & VC   & F    & $r_{\rm sink}$ & $s$       & AT    & ET   & $\alpha$ &total range \\ \hline
lag [orbits]& 0.0           & 0.0  & 0.0        & 0.0  & 0.0  & -0.02-0.0      & 0.0       & 0.0               & 0.0  & 0.0      & $\mathbf{ [-0.02, 0.02) }$     \\ \hline
correlation & 0.88-0.91     & 0.88 & 0.83       & 0.83 & 0.87 & 0.75-0.80      & 0.82-0.89 & 0.84              & 0.81 & 0.85     & {\bf 0.75-0.91 }     \\ \hline
\end{tabular}
\caption{Circular binary: lags and correlations between infrared and optical. A positive lag corresponds to infrared lagging optical.}\label{tab:circ_lag}
\end{table*}

\begin{table*}[t]
\begin{tabular}{|l|l|l|l|l|l|l|l|l|l|l|l||l|}
\hline
            & $\mathcal{M}$ & $D$ & $\Delta x$ & VC   & F    & $r_{\rm sink}$ & $s$       & AT     & ET   & $\alpha$ &$e=0.7$ & total range \\ \hline
lag [orbits]& 0.0-0.08      & 0.0 & 0.0       & 0.0  & 0.0  & 0.02       & -0.02-0.0 & 0.0                & 0.0  & 0.0      & 0.0    & $\mathbf{[-0.02, 0.08]}$     \\ \hline
correlation & 0.80-0.94     & 0.64    & 0.89   & 0.94 & 0.93 & 0.89-0.94      & 0.93-0.96 & 0.78               & 0.94 & 0.88     & 0.90   & {\bf 0.64-0.96 }      \\ \hline
\end{tabular}
\caption{Eccentric binary: lags and correlations between infrared and optical. A positive lag corresponds to infrared lagging optical. All columns are for $e=0.45$ except the column labeled $e=0.7$.}\label{tab:ecc_lag}
\end{table*}


\begin{table*}[t]
\begin{tabular}{|l|l|l|l|l|l|l|l|l|l|l||l|}
\hline
                             & $\mathcal{M}$ & $D$ & $\Delta x$ & VC      & F  & $r_{\rm sink}$    & $s$     & AT  & ET  & $\alpha$ & total range  \\ \hline
color average          & 1.0-1.1       & 1.1 & 1.0        & 1.1     & 1.1& 1.2                                            & 1.0-1.2 & 1.1 & 1.1 & 0.97 & {\bf 0.97-1.2 }      \\ \hline
color variability [RMS \%] & 2.5-5.6       & 4.5 & 2.6        & 4.3     & 4.4& 4.3-4.4                                        & 4.3-4.5 & 4.5 & 4.5 & 6.2  & {\bf 2.4-6.2 }      \\ \hline
color (circumsingle disk) & \multicolumn{11}{c|}{0.94}  \\ \hline
lump achromaticity [vs optical, \%] & 46-76       & 81 & 74        & 81     & 77 & 51-71                                        & 75-79 & 83 & 76 & 47  & {\bf 46-83 }      \\ \hline
lump achromaticity [vs infrared, \%] & 80-98       & 99 & 97        & 99     & 98 & 97-98                                        & 97-98 & 98 & 98 & 96  & {\bf 80-99 }      \\ \hline
\end{tabular}
\caption{Circular binary: properties of the infrared-to-optical color. We quote the average and the root-mean-square variability (RMS). The median peak-to-trough difference is roughly $3\times$ the RMS variability. We also provide the color for the corresponding circumsingle disk. The last two rows show the percentage that the ratio of lump-to-orbital peak frequencies decreased from the optical \& infrared bands to the color. This measures the degree of achromaticity of the lump period, relative to each band.} \label{tab:circ_color}
\end{table*}

\begin{table*}[t]
\begin{tabular}{|l|l|l|l|l|l|l|l|l|l|l|l||l|}
\hline
                             & $\mathcal{M}$ & $D$ & $\Delta x$ & VC      & F  & $r_{\rm sink}$    & $s$     & AT  & ET  & $\alpha$ & $e=0.7$ & total range  \\ \hline
color average          & 1.3-2.0       & 1.5 & 1.7        & 1.5     & 2.0& 1.8-2.1                                        & 1.6-1.7 & 1.7 & 1.7 & 1.7      & 1.5       & {\bf 1.3-2.1 }      \\ \hline
color variability [RMS \%] & 7.2-12        & 6.3 & 7.5        & 9.8     & 12 & 8.2-8.8                                        & 9.9-12  & 9.5 & 9.4 & 12       & 10        & {\bf 6.3-12 }      \\ \hline
color (circumsingle disk) & \multicolumn{12}{c|}{0.94}  \\ \hline
\end{tabular}
\caption{Eccentric binary: properties of the infrared-to-optical color. We quote the average and the root-mean-square variability (RMS). The median peak-to-trough difference is roughly $3\times$ the RMS variability. We also provide the color for the corresponding circumsingle disk. All columns are for $e=0.45$ except the column labeled $e=0.7$.} \label{tab:ecc_color}
\end{table*}
\FloatBarrier

\section{Doppler brightening} \label{app:dop}

Doppler brightening is a relativistic effect appearing first at order $v_\parallel/c$, where $v_\parallel$ is the net line-of-sight velocity of the minidisk. At a given photon frequency $\nu$, assuming that the intrinsic emission is a power-law in $\nu$, $F_\nu\propto \nu^\alpha$, the Doppler brightening signal at lowest order in $v_\parallel/c$ modifies the unboosted flux $F_\nu$ to an observed flux $F^{\rm obs}_\nu$ as $F^{\rm obs}_\nu = F_\nu [1+ (3-\alpha_\nu)v_\parallel/c]$. Here $\alpha_\nu$ is the power spectral index at frequency $\nu$, which is on the order of $\alpha_\nu \simeq -{\rm few}\times 0.1$ \citep[for a very recent and extensive discussion, see][]{Charisi2021}. For the purpose of estimation, we set $\alpha_\nu=-0.44$ \citep[as in][]{berk+2001, Charisi2021} below and replace fluxes with luminosities $F \rightarrow L$.

As a crude approximation, we assume that the net binary motion only Doppler brightens the minidisk emission. To help inform modeling of the Doppler signal, we provide some useful information in Tables~\ref{tab:circ_MD} \&~\ref{tab:ecc_MD}. In particular, expressed as percentages: the average fraction of emission coming from both minidisks ($b$), the average asymmetry of minidisk emission (dimmer minidisk luminosity divided by the brighter one, $f$), and the purely hydrodynamic RMS variability of emission from the minidisks ($\bar{A}_{\rm MD}$) and elsewhere ($\bar{A}_{\rm CBD}$). Note that the median peak-to-trough difference in luminosity (modulated at the orbital frequency) is roughly $3\times$ the RMS variability. 

Let the dimmer minidisk be labeled ``$\,$2,''  with average luminosity $\langle L_2 \rangle \equiv f \langle L_1\rangle$, where $L_1$ is the luminosity of the brighter minidisk. The net luminosity from both minidisks is 
\begin{eqnarray}
    L^{\rm obs}_{\rm MD} &\simeq& L_1 (1+3.44 v_\parallel/c) + L_2 (1 - 3.44 v_\parallel/c),
\end{eqnarray}
\citep[which is a similar to equation (36) from][]{Charisi2021}.
The intrinsic luminosity of the minidisks varies hydrodynamically on the orbital time scale. Let us write this as $L_1 = \langle L_1 \rangle ( 1 + (3/2)A_{1}(t))$, where $\langle L_1\rangle$ is the time-averaged luminosity from minidisk 1 and $A_{1}$ is the fractional RMS hydrodynamic variability. Similarly for minidisk ``2'': $L_2 \simeq \langle L_2 \rangle ( 1 + (3/2)A_{2}(t)) = f\langle L_1 \rangle ( 1 + (3/2)A_{2}(t))$. The line-of-sight velocity also varies on the orbital time scale, with some phase with respect to the hydrodynamic variability; for edge-on observers the effect is maximal, $v_\parallel = v \sin(\Omega_{\rm bin} t)$. The observed luminosity from the minidisks is then
\begin{eqnarray}
    L^{\rm obs}_{\rm MD} &\simeq& \langle L_1 \rangle \left(1+f\right) \left[ 1 + \frac{3}{2} \frac{A_1(t) + f A_2(t)}{1+f} + \left(\frac{1-f}{1+f}\right)3.44\frac{v}{c}\sin(\Omega_{\rm bin}t) \right] + \mathcal{O}\left( \bar{A}_1 \frac{v}{c} \right) + \mathcal{O}\left( \bar{A}_2 \frac{v}{c} \right).
\end{eqnarray}
We neglect the last two terms on the basis that they are of order $\bar{A}_1 (v/c)$ or $\bar{A}_2 (v/c)$. Denote amplitudes with an overbar. We observe in our simulations that $\bar{A}_1 \simeq \bar{A}_2 \equiv \bar{A}_{\rm MD}$, where $\bar{A}_{\rm MD}$ is provided in Tables~\ref{tab:circ_MD} \&~\ref{tab:ecc_MD}. So we approximate $(A_1(t) + f A_2(t))/(1+f) \simeq A_{\rm MD}(t)$.

We must also consider the hydrodynamic variability of non-minidisk emission. Let us call this ``circumbinary'' (CBD) emission, and write $L_{\rm CBD} = \langle L_{\rm CBD}\rangle (1 + (3/2) A_{\rm CBD}(t))$, similarly to the unboosted minidisk emission. $\bar{A}_{\rm CBD}$ is provided (as percentages) in Tables~\ref{tab:circ_MD} \&~\ref{tab:ecc_MD}. The CBD emission has a much weaker Doppler boost, so we treat it as unboosted. We have the total average unboosted emission being $\langle L_{\rm tot} \rangle = \langle L_{\rm CBD}\rangle + \langle L_{\rm MD} \rangle$, and $\langle L_{\rm MD} \rangle = b \langle L_{\rm tot} \rangle$ where $b$ is quoted (as percentages) in Tables~\ref{tab:circ_MD} \&~\ref{tab:ecc_MD}. Then $\langle L_{\rm CBD}\rangle = \langle L_{\rm MD} \rangle (1-b)/b = \langle L_1 \rangle (1+f) (1-b)/b$. The total boosted luminosity is then

\begin{eqnarray}
    L^{\rm obs}_{\rm tot} &\simeq& \langle L_1 \rangle \left(1+f\right) \left[ 1 + \frac{1-b}{b} + \left( \frac{1-b}{b} \right) \frac{3}{2} A_{\rm CBD}(t) + \frac{3}{2} A_{\rm MD}(t) + \left(\frac{1-f}{1+f}\right)3.44\frac{v}{c}\sin(\Omega_{\rm bin}t) \right],
\end{eqnarray}
or rewritten with a prefactor $\langle L_{\rm tot} \rangle = \langle L_1 \rangle (1+f)/b$,
\begin{eqnarray}
    L^{\rm obs}_{\rm tot} &\simeq& \langle L_{\rm tot} \rangle \left[ 1 + \left(1-b\right) \frac{3}{2} A_{\rm CBD}(t) + b \frac{3}{2} A_{\rm MD}(t) + b \left(\frac{1-f}{1+f}\right)3.44\frac{v}{c}\sin(\Omega_{\rm bin}t) \right].
\end{eqnarray}
For clarity, let us define $\delta_{\rm CBD}(t) \equiv (3/2) (1-b) A_{\rm CBD}(t)$, $\delta_{\rm MD}(t) = (3/2) b A_{\rm MD}(t)$, and $\delta_{\rm Dopper} \equiv 3.44 b ((1-f)/(1+f)) (v/c) \sin(\Omega_{\rm bin} t)$, so that we obtain a form of the equation with a similarly simple appearance as Eq.~\eqref{eq:lumschem}:
\begin{eqnarray}
    L^{\rm obs}_{\rm tot} &\simeq& \langle L_{\rm tot} \rangle \left[ 1 + \delta_{\rm CBD}(t) + \delta_{\rm MD}(t) + \delta_{\rm Doppler}(t) \right].
\end{eqnarray}
The CBD and MD fluctuating parts tend to be coherent. To place conservative lower bounds on the orbital velocities which achieve parity between Doppler brightening and hydrodynamic variability, we take the hydrodynamic variability amplitude to be the maximum between the CBD and MD contributions (rather than their coherent sum), i.e. $\bar{\delta}_{\rm hydro} \equiv {\rm max}(\bar{\delta}_{\rm MD}, \bar{\delta}_{\rm CBD})$. Setting this equal to the Doppler variability amplitude $\bar{\delta}_{\rm Doppler}$ and solving for $v/c$, we obtain Eq.~\eqref{eq:dopvmin}:
\begin{eqnarray}
    v/c \simeq \frac{1.5}{3.44\, b} \frac{1+f}{1-f} \max\left( b\bar{A}_{\rm MD}, (1-b) \bar{A}_{\rm CBD} \right).
\end{eqnarray}

\begin{table*}[t]
\begin{tabular}{|l|l|l|l|l|l|l|l|l|l|l||l|}
\hline
              & $\mathcal{M}$ & $D$ & $\Delta x$ & VC & F  & $r_{\rm sink}$ & $s$   & AT & ET & $\alpha$ & total range \\ \hline
$b$ [optical, \%]  & 53-83         & 75  & 81         & 75 & 75 & 78             & 73-77 & 77 & 75 &                    84     & {\bf 53-84 }     \\ \hline
$b$ [infrared, \%] & 15-25         & 21  & 25         & 21 & 21 & 21             & 20-23 & 23 & 21 &                    28     & {\bf 15-28 }     \\ \hline
$f$ [optical, \%]  & 98-100         & 99  & 79        & 99 & 100 & 78-97        & 98-100 & 100 & 100 &                    100     & {\bf 78-100 }     \\ \hline
$f$ [infrared, \%] & 98-100         & 99  & 81        & 99 & 100 & 79-97        & 99-100 & 100 & 100 &                    100     & {\bf 79-100 }     \\ \hline
$\bar{A}_{\rm MD}$ [optical, \%]  & 4.4-12         & 7.8  & 4.5         & 7.5 & 7.7 & 7.1-7.2         & 7.3-8.0 & 7.5 & 7.8 &                    10      & {\bf 4.4-12 }     \\ \hline
$\bar{A}_{\rm MD}$ [infrared, \%] & 3.4-11         & 6.3  & 3.4         & 6.1 & 6.2 & 4.5-5.1         & 5.9-6.2 & 6.1 & 6.3 &                    8.9     & {\bf 3.3-11 }     \\ \hline
$\bar{A}_{\rm CBD}$ [optical, \%]  & 9.7-12        & 10   & 6.2         & 10  & 11  & 5.5-6.7         & 10-11   & 14  & 11  &                    22      & {\bf 5.5-22 }     \\ \hline
$\bar{A}_{\rm CBD}$ [infrared, \%] & 1.2-2.5       & 1.3  & 0.79        & 1.4 & 1.4 & 0.71-0.84       & 1.4     & 1.8 & 1.4 &                    2.5     & {\bf 0.71-2.5 }     \\ \hline
\end{tabular}
\caption{Circular binary. Rows 1 \& 2: average percentage of emission coming from the minidisks (approximated as those regions within $(1/2)\,r_{\rm peri}$ from each black hole). Rows 3 \& 4: the minidisk asymmetry in average luminosity, expressed as a percentage, with the brighter minidisk luminosity always in the denominator (so that the asymmetry is always $\leq100$\%). Rows 5 \& 6: the root-mean-square variability (RMS) averaged over both minidisks (they are similar), expressed as a percentage. Rows 7 \& 8: the RMS variability of the circumbinary disk emission (i.e.~emission from everywhere other than the minidisks), expressed as a percentage. The median peak-to-trough difference is roughly $3\times$ the RMS variability.} \label{tab:circ_MD}
\end{table*}

\begin{table*}[t]
\begin{tabular}{|l|l|l|l|l|l|l|l|l|l|l|l||l|}
\hline
              & $\mathcal{M}$ & $D$ & $\Delta x$ & VC & F  & $r_{\rm sink}$ & $s$    & AT & ET  & $\alpha$ &$e=0.7$ & total range \\ \hline
$b$ [optical, \%]  & 37-73         & 59  & 63         & 64 & 60 & 57-59          & 49-64  & 58 & 55  &                  71     & 43     & {\bf 37-73 }     \\ \hline
$b$ [infrared, \%] & 7.8-17        & 11  & 12         & 13 & 9.5& 9.0-10         & 9.8-12 & 11 & 11  &                  14     & 7.6    & {\bf 7.6-17 }      \\ \hline
$f$ [optical, \%]  & 41-72      & 33 & 52        & 55 & 44 & 35-63      & 52-95 & 55 & 64 &                  45    & 62   & {\bf 33-95 }     \\ \hline
$f$ [infrared, \%] & 47-76      & 39 & 56        & 60 & 50 & 42-67      & 57-94 & 60 & 68 &                  51    & 67   & {\bf 39-94 }      \\ \hline
$\bar{A}_{\rm MD}$ [optical, \%]  & 12-17         & 16  & 11          & 12   & 17 & 12             & 14-18  & 17  & 15  &                  20     & 19      & {\bf 11-20 }     \\ \hline
$\bar{A}_{\rm MD}$ [infrared, \%] & 8.0-12        & 11  & 6.7         & 8.1  & 11 & 6.7-7.4        & 9.1-12 & 11  & 10  &                  15     & 15      & {\bf 6.7-15 }      \\ \hline
$\bar{A}_{\rm CBD}$ [optical, \%]  & 12-23        & 20  & 24          & 29   & 25 & 22-24          & 25-27  & 24  & 20  &                  19     & 21      & {\bf 12-29 }     \\ \hline
$\bar{A}_{\rm CBD}$ [infrared, \%] & 2.1-2.6      & 3.1 & 2.8         & 3.7  & 2.7& 2.3-3.0        & 3.1-3.7& 3.4 & 3.0 &                  1.7    & 3.4     & {\bf 1.7-3.7 }      \\ \hline
\end{tabular}
\caption{Eccentric binary. Rows 1 \& 2: average percentage of emission coming from the minidisks (approximated as those regions within $(1/2)\,r_{\rm peri}$ from each black hole). Rows 3 \& 4: the minidisk asymmetry in average luminosity, expressed as a percentage, with the brighter minidisk luminosity always in the denominator (so that the asymmetry is always $\leq100$\%). Rows 5 \& 6: the root-mean-square variability (RMS) averaged over both minidisks (they are similar), expressed as a percentage. Rows 7 \& 8: the RMS variability of the circumbinary disk emission (i.e.~emission from everywhere other than the minidisks), expressed as a percentage. All columns are for $e=0.45$ except the column labeled $e=0.7$. The median peak-to-trough difference is roughly $3\times$ the RMS variability.} \label{tab:ecc_MD}
\end{table*}
\FloatBarrier

%
%
%
\section{Independent residual test} \label{app:res}

\begin{figure*}[ht]
\centering
\includegraphics[width=1\textwidth]{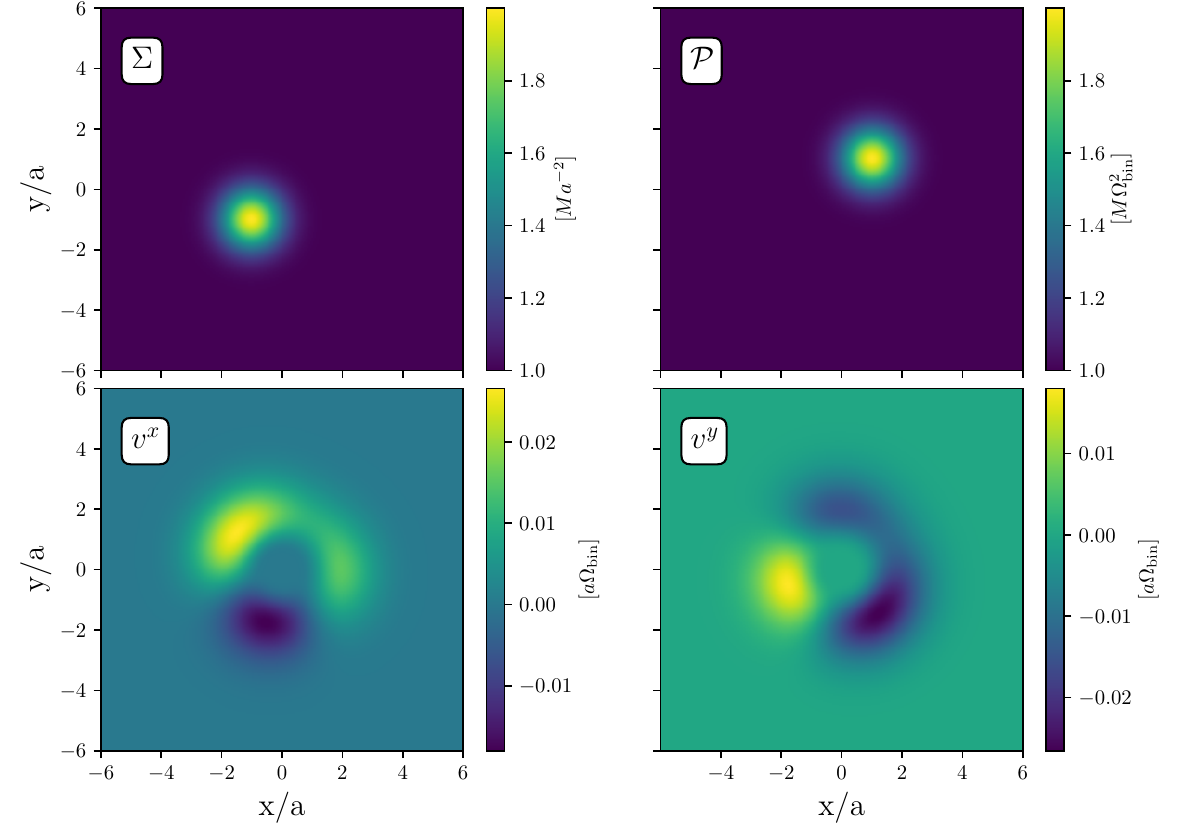}
\caption{Initial conditions for the independent residual test, described in Eqs.~\eqref{eq:indepres}.} \label{fig:resID}
\end{figure*}

\begin{figure*}[ht]
\centering
\includegraphics[width=1\textwidth]{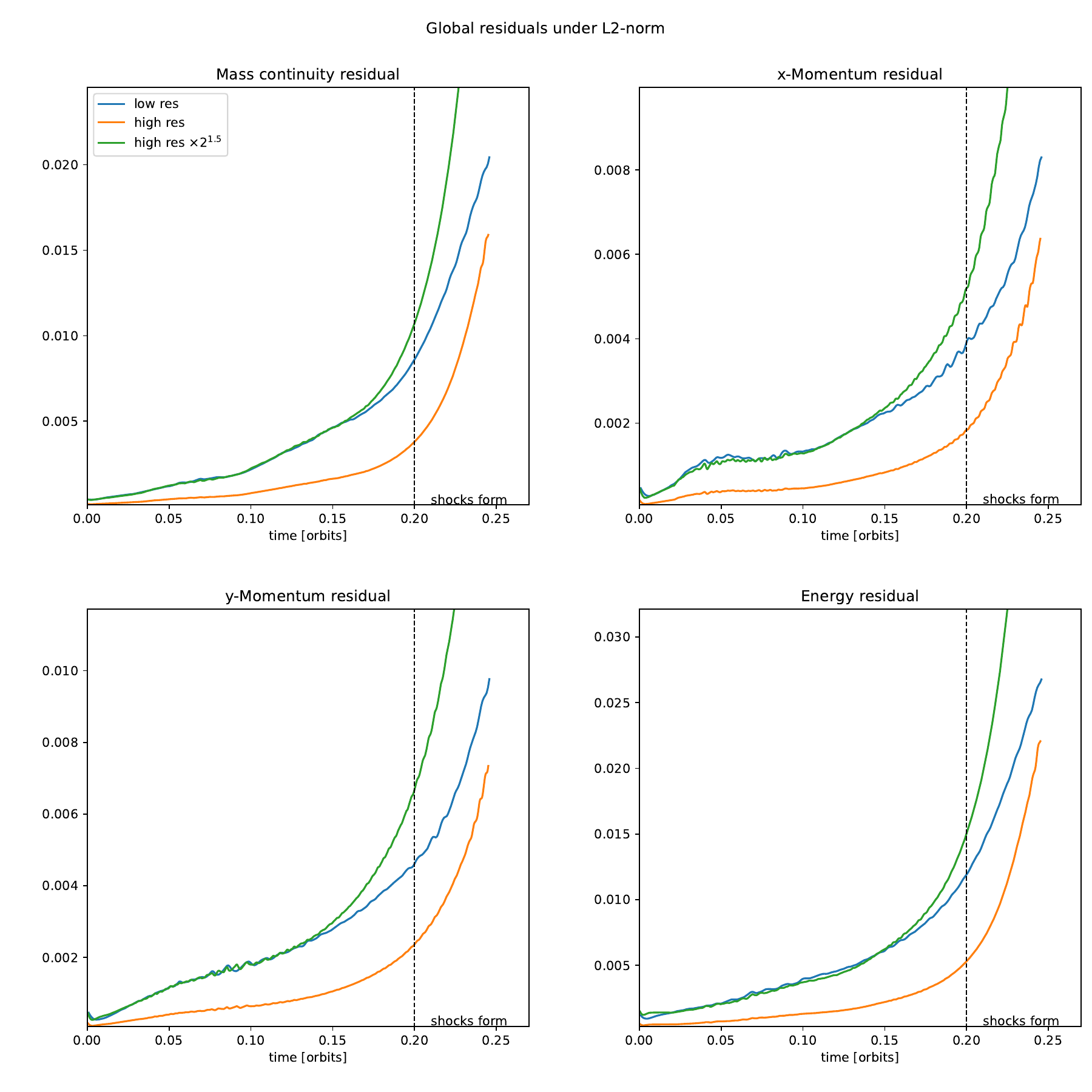}
\caption{Independent residual tests using all the equations of motion. The expected 1.5th-order convergence is obtained while the solution remains smooth, and degrades as shocks form, causing the test validity to break down. The vertical axes are given in units where $M=a=\Omega_{\rm bin}=1$.} \label{fig:res}
\end{figure*}

We test our solution scheme by performing an independent residual test using all the equations of motion. This test consists of plugging a simulated numerical solution into the equations of motion, discretized in space and time using 2nd-order finite differences. The residual evaluator is written independently from the simulation code (hence it is ``independent''), so as not to contaminate it with any bugs that may be present in the simulation code. This is a form of analytic convergence test, since the equations of motion hold analytically. Thus, only two resolutions are necessary for this test. As the accuracy of the numerical solution increases, one expects to see that the equations of motion are being solved to within an error that converges to zero at 1.5th-order. Since we evaluate the equations of motion using finite differences, this test is only valid while the solution is smooth.

The initial conditions for this test are chosen in order to activate all terms in the equations of motion, and such that no symmetries exist. Such conditions ensure a stringent test. The initial conditions we choose are depicted in \fref{fig:resID}, and are given by:
\begin{eqnarray}
    r^2 &\equiv& x^2 + y^2 \nonumber\\
    r_1^2 &\equiv& (x-a)^2 + (y-a)^2 \nonumber\\
    r_2^2 &\equiv& (x+a)^2 + (y+a)^2 \nonumber\\
    \Sigma &=& \left( 1 + \exp{\lbrace -(r_1/a)^2 \rbrace} \right) M a^{-2} \label{eq:indepres}\\
    \mathcal{P} &=& \left( 1 + \exp{\lbrace -(r_2/a)^2 \rbrace} \right) M \Omega_{\rm bin}^{2} \nonumber \\
    v^r &=& \sin \lbrace \phi - \pi/4 \rbrace \exp \lbrace -5(a/r) - (1/3)(r/a)^2\rbrace M \Omega_{\rm bin} \nonumber\\
    v^\phi &=& \sqrt{a/r} \exp \lbrace -5(a/r) - (1/3)(r/a)^2\rbrace M \Omega_{\rm bin}. \nonumber
\end{eqnarray}
We also set $\Gamma=5/3$, $q=0.5$, $M=0.033 M_\odot$, $a=10^{-4}\,$ pc, $\alpha=0.001$, $r_{\rm sink}=a$, $s=0.05$, and $D=6a$. Low resolution corresponds to $\Delta x = 2D/256$ and high resolution corresponds to $\Delta x = 2D/512$. All terms in the equations of motion are active, including radiative cooling (\sref{sec:cooling}), torque-free sinks (\sref{sec:sinks}), and the buffer source terms (\sref{sec:numerics}). The residuals are plotted versus time in \fref{fig:res}; the residuals have been subjected to spatial $\mathcal{L}_2$-norms. The high-resolution residual is also shown scaled up by a factor of $2^{1.5}$, which is expected to coincide with the low-resolution residual while the solution is smooth. The formation of steepening gradients is evident starting around $t\simeq 0.15$ orbits.

\end{document}